\documentstyle[12pt]{article}
\def\hybrid{\topmargin 0pt      \oddsidemargin 0pt
        \parskip 5pt plus 1pt   \jot = 1.5ex}
\catcode`\@=11
\def\marginnote#1{}

\newcount\hour
\newcount\minute
\newtoks\amorpm
\hour=\time\divide\hour by60
\minute=\time{\multiply\hour by60 \global\advance\minute by-\hour}
\edef\standardtime{{\ifnum\hour<12 \global\amorpm={am}%
        \else\global\amorpm={pm}\advance\hour by-12 \fi
        \ifnum\hour=0 \hour=12 \fi
        \number\hour:\ifnum\minute<10 0\fi\number\minute\the\amorpm}}
\edef\militarytime{\number\hour:\ifnum\minute<10 0\fi\number\minute}

\def\draftlabel#1{{\@bsphack\if@filesw {\let\thepage\relax
   \xdef\@gtempa{\write\@auxout{\string
      \newlabel{#1}{{\@currentlabel}{\thepage}}}}}\@gtempa
   \if@nobreak \ifvmode\nobreak\fi\fi\fi\@esphack}
        \gdef\@eqnlabel{#1}}
\def\@eqnlabel{}
\def\@vacuum{}
\def\draftmarginnote#1{\marginpar{\raggedright\scriptsize\tt#1}}

\def\draft{\oddsidemargin -0.1truein
        \def\@oddfoot{\sl preliminary draft \hfil
        \rm\thepage\hfil\sl\today\quad\militarytime}
        \let\@evenfoot\@oddfoot \overfullrule 3pt
        \let\label=\draftlabel
        \let\marginnote=\draftmarginnote
   \def\@eqnnum{{\rm (\theequation)}\rlap{\kern\marginparsep\tt\@eqnlabel}%
\global\let\@eqnlabel\@vacuum}  }

\newdimen\linethick  \linethick=0.4pt
\newdimen\hboxitspace    \hboxitspace=5pt
\newdimen\vboxitspace    \vboxitspace=5pt

\def\fr#1{%
\beq\new
\vcenter{
\hrule height\linethick
           \hbox{\vrule width\linethick
                 \kern\hboxitspace
                 \vbox{\kern\vboxitspace
                       \hbox{$\begin{array}{c}\displaystyle#1
          \end{array}$}%
                       \kern\vboxitspace}%
                 \kern\hboxitspace
                 \vrule width\linethick}%
           \hrule height\linethick}%
\eeq}

\newdimen\Squaresize \Squaresize=14pt
\newdimen\Thickness \Thickness=0.5pt

\def\Square#1{\hbox{\vrule width \Thickness
   \vbox to \Squaresize{\hrule height \Thickness\vss
      \hbox to \Squaresize{\hss#1\hss}
   \vss\hrule height\Thickness}
\unskip\vrule width \Thickness}
\kern-\Thickness}

\def\Vsquare#1{\vbox{\Square{$#1$}}\kern-\Thickness}

\def\numberbysection{\@addtoreset{equation}{section}
        \def\theequation{\thesection.\arabic{equation}}}
\numberbysection

\renewcommand{\theequation}{\thesection.\arabic{equation}}
\newcommand{\l@qq}[2]{\addvspace{2em}
 \hbox to\textwidth{\hspace{1em}\bf #1 \dotfill #2}}

\newcounter{app}

\def\app{\setcounter{equation}{0}
\def\theequation{\Alph{app}.\arabic{equation}}\par
   \addvspace{4ex}
   \@afterindentfalse
  \secdef\@app\@dapp}
\newcommand\@app{\@startsection {app}{1}{0ex}%
                                   {-3.5ex \@plus -1ex \@minus -.2ex}%
                                   {2.3ex \@plus.2ex}%
                                   {\normalfont\Large\bf}}

\def\@dapp#1{%
{\parindent \z@ \raggedright  \bf #1}\par\nobreak}
\def\l@app#1#2{\ifnum \c@tocdepth >\z@
    \addpenalty\@secpenalty
    \addvspace{1.0em \@plus\p@}%
    \setlength\@tempdima{2.5em}%
    \begingroup
      \parindent \z@ \rightskip \@pnumwidth
      \parfillskip -\@pnumwidth
      \leavevmode \bfseries
      \advance\leftskip\@tempdima
      \hskip -\leftskip
      #1\nobreak\hfil \nobreak\hb@xt@\@pnumwidth{\hss #2}\par
    \endgroup\fi}
\newcounter{sapp}[app]

\def\sapp{\def\theequation{\Alph{app}.\arabic{equation}}\par
   \@afterindentfalse
  \secdef\@sapp\@dsapp}
\newcommand\@sapp{\@startsection{sapp}{2}{\z@}%
                                     {-3.25ex\@plus -1ex \@minus -.2ex}%
                                     {1.5ex \@plus .2ex}%
                                     {\normalfont\large\bfseries}}

\def\@dsapp#1{%
{\parindent \z@ \raggedright  \bf #1}\par\nobreak}
\newcommand{\l@sapp}{\@dottedtocline{2}{1.5em}{3em}}

\def\titlepage{\@restonecolfalse\if@twocolumn\@restonecoltrue\onecolumn
     \else \newpage \fi \thispagestyle{empty}\c@page\z@
        \def\thefootnote{\fnsymbol{footnote}} }

\def\endtitlepage{\if@restonecol\twocolumn \else  \fi
        \def\thefootnote{\arabic{footnote}}
        \setcounter{footnote}{0}}  
\relax

\hybrid

\newtoks\@stequation

\def\subequations{\refstepcounter{equation}%
  \edef\@savedequation{\the\c@equation}%
  \@stequation=\expandafter{\theequation}
  \edef\@savedtheequation{\the\@stequation}
  \edef\oldtheequation{\theequation}%
  \setcounter{equation}{0}%
  \def\theequation{\oldtheequation\alph{equation}}}

\def\endsubequations{%
  \setcounter{equation}{\@savedequation}%
  \@stequation=\expandafter{\@savedtheequation}%
  \edef\theequation{\the\@stequation}%
  \global\@ignoretrue}

\makeatletter
\newdimen\normalarrayskip              
\newdimen\minarrayskip                 
\normalarrayskip\baselineskip
\minarrayskip\jot
\newif\ifold             \oldtrue            \def\new{\oldfalse}
\def\arraymode{\ifold\relax\else\displaystyle\fi} 
\def\eqnumphantom{\phantom{(\theequation)}}     
\def\@arrayskip{\ifold\baselineskip\z@\lineskip\z@
     \else
     \baselineskip\minarrayskip\lineskip1\baselineskip\fi}

\def\@arrayclassz{\ifcase \@lastchclass \@acolampacol \or
\@ampacol \or \or \or \@addamp \or
   \@acolampacol \or \@firstampfalse \@acol \fi
\edef\@preamble{\@preamble
  \ifcase \@chnum
     \hfil$\relax\arraymode\@sharp$\hfil
     \or $\relax\arraymode\@sharp$\hfil
     \or \hfil$\relax\arraymode\@sharp$\fi}}

\def\@array[#1]#2{\setbox\@arstrutbox=\hbox{\vrule
     height\arraystretch \ht\strutbox
     depth\arraystretch \dp\strutbox
     width\z@}\@mkpream{#2}\edef\@preamble{\halign \noexpand\@halignto
\bgroup \tabskip\z@ \@arstrut \@preamble \tabskip\z@ \cr}%
\let\@startpbox\@@startpbox \let\@endpbox\@@endpbox
  \if #1t\vtop \else \if#1b\vbox \else \vcenter \fi\fi
  \bgroup \let\par\relax
  \let\@sharp##\let\protect\relax
  \@arrayskip\@preamble}
\def\eqnarray{\stepcounter{equation}%
              \let\@currentlabel=\theequation
              \global\@eqnswtrue
              \global\@eqcnt\z@
              \tabskip\@centering                      
              \let\\=\@eqncr
              $$%
            \halign to \displaywidth  \bgroup
             \eqnumphantom \@eqnsel
      \hskip\@centering                               
    $\displaystyle  \tabskip\z@ {##}$%
    &\global\@eqcnt\@ne \hskip 2\arraycolsep
         $ \displaystyle  \arraymode{##}$\hfil
    &\global\@eqcnt\tw@ \hskip 2\arraycolsep
         $\displaystyle\tabskip\z@{##}$\hfil
         \tabskip\@centering
    &{##}\tabskip\z@\cr}
\makeatother

\def\bea{\begin{eqnarray}}
\def\eea{\end{eqnarray}}

\def\beq{\begin{equation}}
\def\eeq{\end{equation}}
\def\be{\beq\new\begin{array}{c}}
\def\ee{\end{array}\eeq}
\def\bse{\begin{subequations}}                
\def\ese{\end{subequations}}                 %

\begin{document}
\vspace{0.2cm}
\begin{center}
{\LARGE \bf Smooth Gauge Strings and } \\
\vspace{0.5cm}{\LARGE \bf $D\geq{2}$ Lattice Yang-Mills Theories. } \\
\vspace{0.5cm} {\large\bf Andrey Yu. Dubin}\\
\vspace{0.3cm}{\bf ITEP, B.Cheremushkinskaya 25, Moscow 117259, Russia}\\
\vspace{0.3cm}{{\it e-mail: dubin@vxitep.itep.ru}}\\
{tel.:7 095 129 9674  ;  fax:7 095 883 9601}
 \end{center}
\vspace{0.3cm}

\begin{abstract}

Employing the nonabelian duality transformation \cite{Dub2},
I derive the Gauge String form of certain $D\geq{3}$
lattice Yang-Mills ($YM_{D}$) theories in the strong coupling ({\it SC})
phase. With the judicious
choice of the actions, in $D\geq{3}$ our construction generalizes the
Gross-Taylor stringy reformulation of the {\it continuous} $YM_{2}$ on a $2d$
manifold. Using the Eguchi-Kawai model as an example,
we develope the algorithm to determine the weights $w[\tilde{M}]$ for 
connected {\it YM}-flux worldsheets $\tilde{M}$ immersed
into the $2d$ skeleton of a $D\geq{3}$ base-lattice.
Owing to the invariance of $w[\tilde{M}]$ under a {\it continuous} group of
area-preserving worldsheet homeomorphisms,
the set of the weights $\{w[\tilde{M}]\}$ can be used to define
the theory of the {\it smooth} {\it YM}-fluxes which
unambiguously refers to a particular continuous $YM_{D}$ system.
I argue that the latter $YM_{D}$ models (with a {\it finite} ultraviolet
cut-off) for {\it sufficiently large} bare
coupling constant(s) are reproduced, to all orders in
$1/N$, by the smooth Gauge String thus associated.
The asserted $YM_{D}/String$ duality allows to make
a concrete prediction for the 'bare' string tension $\sigma_{0}$
which implies that (in the large $N$ {\it SC} regime) the continuous $YM_{D}$
systems exhibit confinement for $D\geq{2}$. The resulting pattern is 
qualitatively consistent (in the extreme $D=4$ {\it SC} limit)
with the Witten's proposal motivated by the {\it AdS/CFT}
correspondence.

\end{abstract}

\begin{center}
\vspace{0.3cm}{Keywords: Lattice, Yang-Mills, Duality, String} \\
\vspace{0.2cm}{PACS codes 11.15.Ha; 11.15.Pg; 11.15.Me; 12.38.Aw}
\end{center}

\newpage

\section{Introduction}

The quest for a string representation of the $D=4$ continuous Yang-Mills
($YM_{D}$) gauge theory shaped, to a large extent, many branches of the
contemporary mathematical physics. Currently, there are two (qualitatively
overlapping) candidates for the stringy systems conjecturaly dual to 
$YM_{D}$. The first one, due to Polyakov, puts forward certain
Ansatz \cite{Polyak2} for the world-sheet action which is to ensure the
invariance of the
Wilson loop averages $<W_{C}>$ with respect to the zig-zag
{\it backtrackings} of the contour $C$. The complementary approach has
recently sprung to life after daring conjecture of Maldacena \cite{Maldac}
(further elaborated in \cite{PGK,Witt1}) about the so-called {\it AdF/CFT}
correspondence concerning the ${\mathcal{N}}=4$ SUSY $YM_{4}$.
Being motivated by the latter correspondence, Witten has made
a speculative proposal \cite{Witt2} advocating that the ordinary
nonsupersymmetric $YM_{4}$ exhibits confinement at least when considered for
large $N$ in the specific strong coupling ({\it SC}) regime. Namely,
employing certain SUSY $YM_{\bar{D}},~\bar{D}>D,$ properly 
broken to a nonsupersymmetric $YM_{D}$ system, one is to fix an effective {\it finite} 
ultraviolet
({\it UV}) cut off $\Lambda$ while keeping the relevant $YM_{D}$ coupling
constant(s) {\it sufficiently large}.

At present, the support of the conjectured duality-mappings
is fairly limited to a few indirect though reasonably compelling arguments.
On the other hand, final justification of that or another stringy
representation calls either to reproduce the $YM_{D}$ loop-equations
in stringy terms or to provide with
an explicit transformation of the gauge theory into a sort of string theory.
It is our goal to make a step in the second direction approaching (with new
tools) the old challenge: the exact reformulation of the continuous
$D\geq{3}$ $YM_{D}$ theory in terms of the {\it microscopic} colour-electric
{\it YM}-fluxes. More specifically, we focus on the pattern of the
smooth Gauge String inherent in the presumable large $N$ {\it SC} expansion
(as {\it opposed} to the standard weak coupling ({\it WC}) series) valid in
the {\it SC}
regime similar to that of the Witten's proposal \cite{Witt2}. Complementary,
the considered generic strongly coupled continuous $YM_{D}$ models in $D=4$
might be viewed as the prototypes of the effective low-energy
$YM_{4}$ theory. The latter is supposed to result via the
Wilsonian renormgroup ({\it RG}) flow of the {\it effective} actions
(starting from the
standard $YM_{4}$ in the {\it WC} phase) up to the confinement scale.
In this perspective, the 'built in' {\it UV} cut off $\Lambda$ in $D=4$ is
to be qualitatively identified with a physical scale $\Lambda_{YM_{4}}$ that
is of order of the lowest glueball mass.

The short-cut way to a continuous model of smooth {\it YM}-fluxes is
suggested by the two-dimensional analysis. Here, the pattern of the 
flux-theory \cite{Gr&Tayl} proposed by Gross and Taylor is encoded not only
in the continuous $YM_{2}$ systems on a $2d$ manifold. It is also inherent
in the associated {\it RG} invariant $2d$ {\it lattice} gauge models
\cite{Migd75,Witt91} introduced via the plaquette-factor 
\be
Z(\{\tilde{b}_{k}\}|U)=\sum_{R} e^{-F(R)}\chi_{R}(U)~~;~~
e^{-F(R)}=dimR~e^{-\Gamma(\{\tilde{b}_{k}\},N,\{C_{p}(R)\})},
\label{4.1}
\ee
where $dimR,~\chi_{R}(U),$ and $C_{p}(R),~p=1,...,N,$ stand respectively for
the dimension, character  and $p$th order Casimir operator associated to a
given $SU(N)$ irreducible representation (irrep) $R$ (while
$\{\tilde{b}_{k}\sim{N^{0}}\}$ denotes a set of the dimensionless
coupling constants).
The key observation is that, in the $D\geq{3}$ lattice $YM_{D}$ systems
(\ref{4.1}), the pattern of the appropriately constructed 
flux-theory as well refers (owing to a subtle $D\geq{3}$ 'descedant' of the
$2d$ {\it RG} invariance) to the properly associated {\it continuous}
$YM_{D}$
models.

To support this assertion, I first reformulate
the strongly coupled $D\geq{3}$ lattice $YM_{D}$ models (\ref{4.1}) in terms
of the Gauge String which does appropriately extend the $D=2$ Gross-Taylor
stringy pattern \cite{Gr&Tayl} into higher dimensions. This 
lattice theory of the {\it YM}-flux is endowed with certain {\it continuous}
(rather than {\it discrete} as one might expect from the lattice formulation)
group of the area-preserving homeomorphisms.
In turn, the latter symmetry ensures that the considered
$D\geq{2}$ pattern of the lattice flux-theory can be employed to
{\it unambiguously} define the associated $D\geq{2}$ {\it smooth} Gauge
String invariant under the area-preserving diffeomorphisms. 
The remarkable thing is that, in the latter continuous $D\geq{3}$ flux-theory,
one can identify (see Section 8) such {\it SC} conglomerates of the (piecewise) smooth
flux-worldsheets  
which are in one-to-one correspondence with the judiciously associated 
varieties of the {\it WC} Feynman diagrams
on the side of the properly specified  continuous $YM_{D}$ model.
Therefore, the proposed smooth Gauge String provides with the concrete
realization of the old expectation \cite{'tHooft1} (for a recent discussion
see \cite{Witt2}) that certain nonperturbative effects 'close up' the windows
of the Feynman diagrams trading the latter for the string worldsheets.

More specifically, building on the nonabelian duality transformation
\cite{Dub2} recently proposed by the author, we show that in the $D\geq{2}$
lattice systems (\ref{4.1}) the free energy and the Wilson (multi)loop
observables can be rewritten in terms of the following statistics of strings.
The lattice weight $w[\tilde{M}(T)]$ of a given connected worldsheet
$\tilde{M}(T)$ (with the support on a subspace, represented by certain $2d$
cell-complex {\it T}, of the $2d$ skeleton of the $D$-dimensional
base-lattice)
\be
w[\tilde{M}(T)]=exp[{-\bar{A}\Lambda^{2}\tilde{\sigma}_{0}(\{\tilde{b}_{k}\})}~]~
N^{2-2h-b}~J[\tilde{M}(T)|\{\tilde{b}_{k}\}]~,
\label{0.5c}
\ee
is composed of the three different blocks which altogether conspire so
that the contribution of the strings with any {\it backtrackings} (i.e.
foldings) is zero. In eq.
(\ref{0.5c}), the first factor is the exponent of the Nambu-Goto term
proportional to the total area $\bar{A}$ of $\tilde{M}$. The {\it D-independent}
bare string tension ${\sigma}_{0}=\Lambda^{2}\tilde{\sigma}_{0}$ (to be
defined by eq. (\ref{0.1ef}) below) is measured in the units of the {\it UV}
cut off $\Lambda$ squared (which, in $D\geq{3}$, regularizes the 'transverse'
string fluctuations). Next, there appears the standard 't Hooft factor, where
{\it h} and {\it b} are respectively the genus and the number of the
boundary contours $C_{k},~k=1,...,b,$ of
$\tilde{M}$.

Finally, given a particular model (\ref{4.1}), the third term
$J[\tilde{M}(T)|\{\tilde{b}_{k}\}]$ (being equal to {\it unity} for a
nonselfintersecting surface $\tilde{M}$) is sensitive only to the
{\it topology} (but not to the geometry) of selfintersections of $\tilde{M}$.
In particular, the dependence of
$J[\tilde{M}(T)|\{\tilde{b}_{k}\}]$ on the coupling constants
$\{\tilde{b}_{k}\}$ is collected from the elementary weights
assigned to the admissible 'movable' singularities (to be specified later
on) of the associated map $\varphi:~\tilde{M}\rightarrow{T}$.
 As a result, the last term (similarly to the remaining ones) is 
invariant under the required continuous group of the area-preserving
homeomorphisms so that the pattern (\ref{0.5c}) directly applicable to a
{\it generic} smooth worldsheet $\tilde{M}$.

Next, consider the $D\geq{3}$ continuous flux-theory defined as the statistics
of the smooth worldsheets $\tilde{M}$ which are postulated to be endowed with
the weights (\ref{0.5c}) corresponding to the smooth mappings of $\tilde{M}$
into the Euclidean space ${\bf R^{D}}$. In compliance with the above
asserted {\it SC}/{\it WC}
correspondence, a given specification (\ref{4.1}) the smooth Gauge String
refers to the following {\it unique} continuous $YM_{D}$ model. The {\it
local}
lagrangian of the latter is to be reconstructed as the $D$-dimensional
'pull-back'
\be
L_{2}(F)\rightarrow{L_{D}(F)}~:~\{F^{a}_{\mu\nu}~;~\mu,\nu=1,2\}
\longrightarrow{\{F^{a}_{\mu\nu}~;~\mu,\nu=1,...,D\}}
\label{0.1ev}
\ee
of the $D=2$ lagrangian (composed of the $O(D)$-invariant combinations
of the $F^{a}_{\mu\nu}\equiv{F}$ tensor)
\be
D=2~~~:~~~L_{D}(F)=\Lambda^{D-4}\sum_{n\geq{2}}~\sum_{r\in{Y_{n}}}
\frac{[tr(F^{k})]^{p_{k}}}{g_{r}(\{\tilde{b}_{k}\})}~,
\label{0.1ee}
\ee
associated to such continuous $YM_{2}$ theory \cite{Witt91,Dougl&Li} that
its partition function on a $2d$ disc (of unit area with the free
boundary conditions) is equal to the plaquette-factor (\ref{4.1}).
(In $D\geq{3}$, within each trace of (\ref{0.1ee}), the involved
$(F_{\mu\nu})^{i}_{j}$-factors are prescribed to be totally symmetrized to
exclude the $[F_{\mu\nu},F_{\rho\sigma}]$-commutator dependent
terms identically vanishing in $D=2$.)
To separate out the 'kinematical' {\it D}-{\it dependent} rescaling of the 
coupling constants, we have introduced the parameter $\Lambda$. The
latter is to be identified with the effective {\it UV} cut off, for 
the considered strongly coupled $YM_{D}$ systems, predetermined by
the necessary regularization of the $D\geq{3}$ {\it YM}-flux transverse
fluctuations. Remark also that the {\it N}- and $\Lambda$-dependent coupling
constants ${g}_{r}(\{\tilde{b}_{k}\})
\equiv{{g}_{r}(\{\tilde{b}_{k}\},N,\Lambda)}$ are canonically
labelled by the $S(n)$ irreps $r\in{Y_{n}}$ (parametrized by the
partitions of $n$: $\sum_{k=1}^{n} kp_{k}=n$).

In sum, our proposal is that
{\bf the particular theory of the smooth Gauge String, thus induced
through (\ref{4.1}), reproduces (to all orders in $1/N$) the corresponding
local continuous $D\geq{3}$ $YM_{D}$ model (\ref{0.1ee}) for sufficiently
large coupling constants $\{g_{r}N^{2-\sum_{k=1}^{n}
p_{k}}\sim{O(N^{0})}\}$}. (The latter constants are
measured in the units of $1/N$ and $\Lambda$ akin to
eq. (\ref{7.2n}).)
Note that in $D=2$, after proper
identification of the
coupling constants, this correspondence is justified by the matching
with the $2d$ pattern \cite{Gr&Tayl} of Gross and Taylor.
One may also expect that the asserted $YM_{D}/String$ correspondence is unique
to the extent that the stringy degrees of freedom are {\it directly}
identified with the worldsheets of the {\it microscopic, conserved}
{\it YM}-flux immersed (by mappings with certain admissible singularities to
be specified after eq. (\ref{0.1}) below) into a $D$-dimensional base-space.

The above constriant on $\{g_{r}\}$ (selecting the $SC$ regime) accounts for
the fact that in $D\geq{3}$ the proposed $YM_{D}/String$ duality is
limited by the stability of the {\it YM}-flux. In
particular, the physical string tension $\sigma_{ph}$ (entering e.g. the
asymptotics of the Wilson loop averages) must be positive. As we
will see, despite the extra $J[..]$-factor in eq. (\ref{0.5c}), the 'bare' and
the physical string tensions (defined within the
$1/N$ expansion) are conventionally related so that the latter
condition in the large $N$ limit reads
\be
\{~{g}_{r}(\{\tilde{b}_{k}\},..)~\}~:~~~\sigma_{ph}=
(\tilde{\sigma}_{0}(\{\tilde{b}_{k}\})-\zeta_{D})\cdot\Lambda^{2}~>0~,
\label{0.7b}
\ee
i.e. the 'bare' part $\sigma_{0}=\Lambda^{2}\tilde{\sigma}_{0}$ should be
larger than the $D\geq{3}$ entropy contribution $\delta \sigma_{ent}=
-\zeta_{D}\Lambda^{2}$ due to the transverse string fluctuations. Among our
results, the central one is the explicit
$\{\tilde{b}_{k}\}$-dependence of 
$\sigma_{0}=\Lambda^{2}\tilde{\sigma}_{0}(\{\tilde{b}_{k}\})$ entering the
worldsheet weights (\ref{0.5c}). It is uniquely reconstructed
\be
\Gamma(\{\tilde{b}_{k}\},N,\{C_{p}(R)\})=n\tilde{\sigma}_{0}(\{\tilde{b}_{k}\})
(1+O(N^{-1}))~~~;~~~~
\sigma_{0}=\Lambda^{2}\tilde{\sigma}_{0}\sim{O(N^{0})},
\label{0.1ef}
\ee
from the formal $1/N$ asymptotics (with $R\in{Y_{n}^{(N)}},~
n\sim{O(N^{0})}$) of the admissible function $\Gamma(..)$ defining the
associated lattice model (\ref{4.1}). For example, the smooth Gauge String
induced from the Heat-Kernal action \cite{Dr&Zub}
\be
Z(g^2|U)=\sum_{n\geq{0}}\sum_{R\in{Y_{n}^{(N)}}}
dimR~\chi_{R}(U)~exp[-\tilde{g}^2 C_{2}(R)/2]~,
\label{4.1b}
\ee
being presumably dual to the standard ${\it YM_{D}}$ theory with
${g}^{2}=\Lambda^{4-D}\tilde{g}^{2}\sim{1/N}$,
refers to the following bare string
tension (with $g^{2}\sim{1/N}$)
\be
L_{D}(F)=tr(F^{2}_{\mu\nu})/4g^{2}~~~\Longleftrightarrow{~~~
\sigma_{0}=g^{2}N\Lambda^{D-2}/2}~.
\label{0.1ew}
\ee

Summarizing, the predicted pattern (\ref{0.1ef}) of $\sigma_{0}$ implies that
the continuous $D\geq{3}$ $YM_{D}$ systems are confining at least
in the large $N$ {\it SC} regime belonging to the domain (\ref{0.7b}).
Although in this regime the standard {\it WC} series are expected to
fail, the latter {\it SC} phenomenon suggests a mechanism of confinement
for $YM_{4}$ in the {\it WC} phase (see Conclusions).
Remark also that the physical
string tension in eq. (\ref{0.7b}) generically is {\it not} adjusted to be
infinitely
less than the squared {\it UV} cut off $\Lambda$ (which in $D=4$ matches with
the above identification $\Lambda\sim{\Lambda_{YM_{4}}}$). Actually,
in the considered $YM_{D}$ models the introduced smooth
flux-worlsheets might become unstable degrees of freedom even {\it prior}
to the saturation of (\ref{0.7b}) owing to the
presumable large $N$ phase transition(s) (generalizing the $D=2$ situation 
\cite{Dougl&Kaz}).

Finally, the distinguished regime of the smooth Gauge String is the extreme
large {\it N} {\it SC} limit where $\sigma_{0}(\{\tilde{b}_{k}\})>>
\Lambda^{2}$. As in eq. (\ref{0.7b}) the entropy-constant $\zeta_{D}$ is
$\{\tilde{b}_{k}\}$-independent, in this limit the leading order of the Wilson
loop averages $<W_{C}>$ is given by the minimal area contiribution that allows
for a number of nontrivial predictions. First of all, the
physical string tension in this regime merges
with the 'bare' one, $\sigma_{ph}\rightarrow{\sigma_{0}}$, determined by eq.
(\ref{0.1ef}). (We will show in Section 9 that the Witten's {\it SC}
asymptotics \cite{Witt2} of $\sigma_{ph}$ in $D=4$ is semiquantitatively
consistent with the latter prediction). Also, provided the minimal area
worldsheet $\tilde{M}_{min}(C)$ has the support $T_{min}(C)$ on a
$2d$ manifold (rather than on a $2d$
cell-complex), in this limit the pattern of the $D\geq{3}$ averages $<W_{C}>$
is reduced to the one in the corresponding continuous $D=2$ $YM_{2}$ theory
(\ref{0.1ee}) conventionally defined on $T_{min}(C)$.

\subsection{The $D\geq{2}$ pattern of the Gauge String weights.}
$${\bf A.~~~The~D=2~case.}$$
Recall that, in the $2d$ framework initiated by Gross and Taylor, the
partition function $\tilde{X}_{M}$ of a given continuous $YM_{2}$ theory on a
$2d$ manifold $M$ of the area $A$ (without boundaries) is rewritten as the
sum
\be
\tilde{X}_{M}=\int_{\tilde{M}} ~df~ (-1)^{P_{f}} ~
\frac{N^{\chi_{f}}}{|C_{f}|}~
exp[{-\sigma_{0} n_{f} A}]~~~~;~~~~~f:~\tilde{M}\rightarrow{M},
\label{0.5}
\ee
over the topologically distinct branched covering spaces $\tilde{M}$ of
$M$ specified by the mappings {\it f}. The latter maps locally
satisfy the definition \cite{Massey,StabMap} of the immersion
$\tilde{M}\rightarrow{M}$ everywhere on $M$ except for a set of isolated
points where the singularities, corresponding to
a branch-point or/and to a collapsed (to a point) subsurface connecting a
few sheets, are allowed. To visualize the pattern of $\tilde{M}$
as the associated Riemann surface ({\it without} foldings but with the singularities 
to be viewed as certain collapsed $2d$ subsurfaces) identified with a
particular string worldsheet $\tilde{M}$ of the total Euler characteristic $\chi_{f}$, 
I propose the cutting-gluing rules which readily generalize to the $D\geq{3}$
case. As for $|C_{f}|,P_{f}$, and $n_{f}$, in eq. (\ref{0.5})
they denote respectively the symmetry factor (i.e. the number of distinct
automorphisms $\kappa$ of $f:~f\circ \kappa=f$), the 'parity', and the degree
(i.e. the number of
the covering sheets) of the map $f$. The sum (\ref{0.5}) includes all
admissible disconnected contributions in such a way that the free energy
$-ln[\tilde{X}_{M}]$ is supposed to be deduced \cite{Gr&Tayl} from (\ref{0.5})
constraining the worldsheets $\tilde{M}$ to be connected.

To make contact with the $T=M,~\bar{A}=n_{f}A$ case of (\ref{0.5c}), observe first 
that the summation (\ref{0.5}) over the maps $f$ effectively enumerates the
admissible selfintersections of $\tilde{M}$.
In particular, the sum over the singularities of $f$ 
implies the one over the 'nonmovable' and (the positions of) the 'movable'
branch points parametrized by the corresponding cyclic decomposition
$\{p\}$: $\sum_{k=1}^{n} kp_{k}=n_{f}$ of $n_{f}$.
These two types of the points are respectively assigned with the 
nontrivial weights $\tilde{w}_{\{p\}}$ and $w_{\{p\}}(\{\tilde{b}_{k}\})$
(included into the measure $df$ of (0.5)) which are
$\{\tilde{b}_{k}\}$-dependent only in the
'movable' case. (The $w_{\{p\}}$-, $\tilde{w}_{\{p\}}$-factors, together with
$(-1)^{P_{f}}/|C_{f}|$, are collected into
$J[\tilde{M}(T)|\{\tilde{b}_{k}\}]$ of (\ref{0.5c}).) Therefore, the summation
$df$ implies in particular the multiple integrals
\be
\prod_{\{p\}}~
\frac{[w_{\{p\}}(\{\tilde{b}_{k}\})]^{m_{\{p\}}}}{[m_{\{p\}}]!}~
\prod_{k=1}^{m_{\{p\}}}~\int_{M} d^{2}X^{(k)}_{\{p\}}
\label{0.5cx}
\ee
over all positions $X^{(k)}_{\{p\}}$ on {\it M} of (a given number
$m_{\{p\}}$ of) the movable branch points that introduces additional
dependence on the area $A=\int_{M} d^{2}X^{(k)}_{\{p\}}$ of {\it M}.
('Nonmovable' singularities of the $f$-map can be 
placed anywhere on $M$ but do not carry any area- or 
$\{\tilde{b}_{k}\}$-dependent factors.) Finally,
the bare string tension
$\sigma_{0}=\Lambda^{2}\tilde{\sigma}_{0}(\{\tilde{b}_{k}\})$ (defined via
eq. (\ref{0.1ef})) enters the exponent of (\ref{0.5}) as the
{\it no-fold variant} of the Nambu-Goto term. Our attention is
mostly confined to the option (\ref{0.1ew}) corresponding to the $D\geq{2}$
$SU(N)$ Heat-Kernal lattice gauge theory (\ref{4.1b}). In this case,
${\sigma}_{0}=\Lambda^{2}\tilde{g}^{2}N/2$ and there are only
the simple transposition branch points (i.e. the ones connecting a pair of
the sheets) weighted by $w_{2}=\tilde{g}^{2}N$.

Finally, the $D=2$ representation (\ref{0.5}) of the continuous
$YM_{2}$ theories (\ref{0.1ee}) is valid
also in the corresponding $2d$ {\it lattice} gauge systems (\ref{4.1}) on a
discretized surface $M$.
The origin of this two-fold interpretation of (\ref{0.5}) resides in the
{\it RG} invariance of the latter lattice models. Complementary,
owing to the symmetry of the continuous $YM_{2}$ models (\ref{0.1ee}) under
the area-preserving diffeomorphisms, the {\it RG} invariance of (\ref{4.1})
results in the invariance of the lattice weights
$w[\tilde{M}]$ under the {\it continuous} group of
the area-preserving worldsheet homeomorphisms. These homeomorphisms
{\it continuously} (rather than discretely) translate the
positions of the singularities of the map $f$ everywhere
on $M$ including {\it interiors} of the plaquettes.

$${\bf B.~~~Extension~to~the~D\geq{3}~case.}$$
A given $D\geq{3}$ lattice $YM_{D}$ system
(\ref{4.1}) (having some lattice ${\bf L^{D}}$ as the base-space $B$) can be
equally viewed as the $YM$ model (\ref{4.1}) defined on the $2d$
skeleton ${\bf T^{D}}$ of ${\bf L^{D}}$ represented by the associated
$2d$ cell-complex. As we will discuss in Section 7, this reformulation allows
to reveal certain $D\geq{3}$ 'descedant' of the $2d$ {\it RG} invariance of
(\ref{4.1}) that in turn foreshadows the existence of the $D\geq{3}$ extension
of the pattern (\ref{0.5}).

Consider the partition function $\tilde{X}_{L^{D}}$ of the
$YM_{D}$ system (\ref{4.1}) defined on $B={\bf L^{D}}$.
Combining the nonabelian duality transformation
\cite{Dub2} with the methods of algebraic topology \cite{Massey}, we
derive that the pattern (\ref{0.5}) remains to be valid in $D\geq{3}$ with the
only modification. It concerns the structure of the relevant mappings (to be
summed over) specifying the admissible topology of the worldsheets
$\tilde{M}$ of the {\it YM}-flux. The $D=2$ maps $f$ are superseded by the
mappings $\varphi$ 
\be
\varphi~:~~~\tilde{M}~\longrightarrow{T}~~~\in{~~~{\bf T^{D}}},
\label{0.1}
\ee
of a  $2d$ surface
$\tilde{M}$ onto a given subspace
{\it T} (represented by a $2d$ cell-complex) of the $2d$ skeleton ${\bf
T^{D}}$
of the $D$-dimensional base-lattice ${\bf L^{D}}$. Akin to the $D=2$ case,
the maps $\varphi$ locally comply with the definition \cite{Massey,StabMap}
of the immersion
$\tilde{M}\rightarrow{T}$ anywhere except for a countable set of points on
$T$ where certain singularities are allowed. The relevant (for the smooth
implementation of the Gauge String) ones are either of the {\it same}
type as in the $2d$ $f$-mappings (\ref{0.5}), i.e. the branch points or/and
the collapsed subsurfaces,  or of the type corresponding to the
{\it homotopy retraction} (see e.g. Appendix C) of the latter irregularities.
In particular, the included into $d\varphi$-measure
$\{\tilde{b}_{k}\}$-dependent weights $w_{\{p\}}(\{\tilde{b}_{k}\})$ (of the
movable branch points) in $D\geq{3}$ are equal to their counterparts in the
$2d$ case (\ref{0.5cx}) associated to the same continuous $YM_{2}$ theory
(\ref{0.1ee}). (Akin to eq. (\ref{0.5}), one can argue
that the free energy $-ln[\tilde{X}_{L^{D}}]$ is provided by the restriciton
of the sum for partition function $\tilde{X}_{L^{D}}$ to the one over the
connected worldsheets $\tilde{M}$.)

The $D\geq{3}$ nature of the mappings (\ref{0.1}) is reflected by
the fact that, owing to possible 'higher-dimensional' selfintersections of
$\tilde{M}(T)$,
the taget-space {\it T} generically is represented by the $2d$ cell-complex
$T=\cup_{k} E_{k}$ rather than by a $2d$ surface. (Recall that any $2d$
cell-complex, after cutting
out along the links shared by {\it more} than two plaquettes, becomes
\cite{Dr&Zub,Massey} a disjoint union of $2d$ surfaces $E_{k}$ of the areas $A_{k}$
with certain
boundaries. Conversely, the set $\{E_{k}\}$ can be combined back into
$T=\cup_{k} E_{k}$ according to the incidence numbers \cite{Dr&Zub}
corresponding to the entries of the associated {\it incidence-matrix}.)
Therefore, the construction of $\tilde{M}(T)$
(via our cutting-gluing algorithm developed for the canonical branched
coverings (\ref{0.5})) entails the appropriate generalization of the notion of
the Riemann surface (with the total area 
$\bar{A}=\sum_{k}n^{(k)}_{\varphi}A_{k}$). Alternatively, 
$\tilde{M}(T)$ can be viewed as
the specific  {\it generalization} of the branched covering space (wrapped
around {\it T}) suitable for the analysis of the Gauge String.

Next, as it is shown in Section 8, the total contribution of
the worldsheets $\tilde{M}$ with arbitrary {\it backtrackings} (that,
generalizing the $D=2$ case, bound any zero 3-volume) {\it vanishes}.
The simplest $J[\tilde{M}(T)|\{\tilde{b}_{k}\}]=1$ pattern (\ref{0.5c})
of $w[\tilde{M}]$ in $D\geq{2}$ arises when the genus $h$ connected string
worldsheet $\tilde{M}(T)$ is represented by an {\it embedding}
$T=\tilde{M}(T)$ (i.e. $\tilde{M}(T)$
does {\it not} selfintersect). Nevertheless, the set of all possible
$J[\tilde{M}(T)|\{\tilde{b}_{k}\}]=1$ embedding-weights (\ref{0.5c})
does {\it not} discriminate between those distinct models
(\ref{4.1})/(\ref{0.1ee}) which provide with one and the same bare string
tension (\ref{0.1ef}). To distinguish between the different models, {\it the
specification of the remaining immersion-weights (assigned to the
selfintersecting string worlsheets) is indispensable}. In this way, one
reconstructs the data encoded in a given continuous
$YM_{2}$ theory (\ref{0.1ee}) that originally served as
the bridge (\ref{0.1ev}) between the Gauge String and the corresponding
continuous $YM_{D}$ model.

Owing to the asserted pattern, the $D\geq{3}$ lattice weights
$w[\tilde{M}(T)]$ are invariant under certain continuous group of the
area-preserving homeomorphisms extending the $D=2$ ones. As a
result, the set $\{w[\tilde{M}(T)]\}$ can be unambiguously used to introduce
the statistics of the (piecewise) {\it smooth} {\it YM} flux-worldsheets
$\tilde{M}(T)$, with the latter homeomorphisms being
traded for the corresponding diffeomorphisms. As for the sum over the
worldsheets, it is specified by the one running
over the $\varphi$-mappings (\ref{0.1}). This time, one is to consider the
piecewise smooth immersions (with the admissible singularities which, in $D\geq{4}$,
can be restricted to the ones listed after eq. (\ref{0.1})) of the $2d$
{\it manifolds} $M$ into the $D$-dimensional space-time $B={\bf R^{D}}$ that results 
in the worldsheet $\tilde{M}$ with the support on $T\in{\bf R^{D}}$.
In turn, it constitutes the proper class of the $2d$
cell-complexes {\it T} which can play the role of the taget-spaces
for the considered (piecewise) {\it smooth} maps $\varphi$.

As we will discuss in Section 8, the $D\geq{3}$ dynamics of the smooth Gauge
String in certain sense is dramatically {\it simpler} compared to its lattice
counterpart. First of all, it is 
convenient to make certain redefinition of the bare string tension
(that amounts to the substitution $\tilde{\sigma}_{0}=
\lambda/2\rightarrow{\lambda(1-1/N^{2})/2}$ in the case of 
(\ref{4.1b})) to get rid of a redundant subset of the 'movable' 
singularities present in the $SU(N)$ mappings (\ref{0.1}) even for a 
1-sheet covering 
of $T$. Given this modification, consider the subset of the 
smooth worldsheets $\tilde{M}(T)$ (without boundaries) which are 
constrained to be strictly nonselfintersecting in $D\geq{5}$  and 
allowed  to selfintersect at an arbitrary union of isolated points in
$D=4$. (Actually the latter condition can weakened e.g. to include 
non(self)intersecting boundary contour(s)). They are assigned with the
simplest $J[\tilde{M}(T)|\{\tilde{b}_{k}\}]=1$ weight-pattern (\ref{0.5c}).
The point is that {\bf in $D\geq{4}$ the latter subset is dense in the set of
all worldsheets $\tilde{M}$ parametrized by the (piecewise) smooth mappings
(\ref{0.1})
into ${\bf R^{D}}$}, provided the latter
redefinition of $\tilde{\sigma}_{0}$. In particular, it justifies (at least in $D\geq{4}$) 
the validity of the simple relation (\ref{0.7b}). As for the more complicated
pattern (\ref{0.5c}) of the weights, in $D\geq{4}$ it is observable for
example within such Wilson loop averages $<W_{C}>$ where the corresponding
minimal surface $S_{min}(C)$ selfintersects (or when the boundary contour
$C$ has some zig-zag backtrackings).

\section{Outline of the further content.}

To make the analysis of the $D\geq{2}$ lattice Gauge String more concise,
we employ the Twisted
Eguchi-Kawai (TEK) representation \cite{Gonz-Arr} of the large $N$
'infinite-lattice' {\it SU(N)} gauge systems. Recall that in the limit
$N\rightarrow{\infty}$ the partition function (PF) $\tilde{X}_{ L^{D}}$ of
a lattice $YM_{D}$ theory like (\ref{4.1}) in a $D=2p$ volume
$L^{D}=N^{2}$ can be reproduced (at least within both the {\it SC} and the
{\it WC} series)
\be
\lim_{N\rightarrow{\infty}} \tilde{X}_{ L^{D}}=\lim_{N\rightarrow{\infty}}
\left(\int{\prod_{\rho=1}^{D}{dU_\rho}}~\prod_{\mu\nu=1}^{D(D-1)/2}
{Z(g^2|t\cdot U_{\mu}U_{\nu}U_{\mu}^{+}U_{\nu}^{+})}\right)^{L^{D}},
\label{2.1}
\ee
through the PF $\tilde{X}_{D}$ of the associated $D$-matrix {\it SU(N)} TEK
model with the reduced space-time dependence (where
$t=\exp{[i 2\pi/{N}^{\frac{2}{D}}]}\in{\bf Z_{N}}$). Therefore, in the
$N\rightarrow{\infty}$ {\it SC} phase, the free energy of (\ref{2.1}) yields
the generating functional for the $YM_{D}$
string-weights assigned to the worldsheets corresponding to the $2d$ spheres
immersed into ${\bf L^{D}}$. On the side of the TEK model, the surfaces are
'wrapped around' the EK base-lattice $T_{EK}$ which has the topology of
the $2d$-{\it skeleton} of the $D$-dimensional cube with periodic boundary
conditions. Thus, $T_{EK}$ is homeomorphic to the $2d$ cell-complex 
visualized as the union
\be
T_{EK}=\cup_{\mu\nu=1}^{D(D-1)/2}~E_{\mu\nu}~~~~~,~~~~~
E_{\rho\mu}\cap E_{\rho\nu}=l_{\rho}~,
\label{0.3}
\ee
of the $D(D-1)/2$ mutually intertwined 2-tora $E_{\mu\nu}$ sharing {\it D}
uncontractible cycles (i.e. compactified links of the $D$-cube)
$l_{\rho}$ in common.

To proceed further, let us first label the {\it SU(N)} irreps summed up in
each $\mu\nu$-species of the {\it Z}-factor (\ref{4.1}) entering (\ref{2.1}) by
\be
R_{\mu\nu}\in{Y_{n_{\mu\nu}}^{(N)}}~~~,~~~
n_{+}=\sum\limits_{\mu\nu=1}^{D(D-1)/2} n_{\mu\nu}.
\label{2.1bx}
\ee
It is convenient to rewrite the {\it SU(N)} TEK PF (\ref{2.1}) in the
following form
\be
\tilde{X}_{D}=
\sum_{\{n_{\mu\nu}\}}\sum_{\{R_{\mu\nu}\in{Y_{n_{\mu\nu}}^{(N)}}\}}
t^{n_{+}}e^{-S(\{R_{\mu\nu}\})}B(\{R_{\mu\nu}\})
=\sum_{\{n_{\mu\nu}\}} t^{n_{+}}B(\{n_{\mu\nu}\}),
\label{2.1b}
\ee
introducing the elementary master-integrals $B(\{R_{\mu\nu}\})$
(which are then composed into $B(\{n_{\mu\nu}\})$)
\be
B(\{R_{\mu\nu}\})=\int{\prod_{\{\rho\}}{dU_\rho}}
\prod_{\{\mu\nu\}}
{\chi_{R_{\mu\nu}}(U_{\mu\nu})}~~;~~e^{-S(\{R_{\mu\nu}\})}=
\prod_{\{\mu\nu\}} e^{-F(R_{\mu\nu})},
\label{5.1bc}
\ee
where $U_{\mu\nu}\equiv{U_{\mu}U_{\nu}U_{\mu}^{+}U_{\nu}^{+}}$, $F(R)$ is
defined by eq. (\ref{4.1}), and we have used the
identity $\chi_{R}(tV)=t^{n({R})}\chi_{R}(V),~
t\in{Z_{N}},~R\in{Y_{n({R})}^{(N)}}$. In the {\it SC} phase,
we are concerned below, the twist-factor {\it t} in eq. (\ref{2.1b})
is irrelevant \cite{Gonz-Arr} in the limit $N\rightarrow{\infty}$ so that
the TEK model is reduced to the original Eguchi-Kawai one. In this
regime, the $t=1$ correspondence (\ref{2.1}) is supposed to be valid in any
(not necessarily even) $D\geq{2}$.

Next, making use of the nonabelian duality transformation
\cite{Dub2}, in Section 3 we rewrite the TEK master-integral
$B(\{n_{\mu\nu}\})$ as the weighted sum of the $Tr_{n_{+}}$-characters (i.e.
traces)
\be
B(\{n_{\mu\nu}\})=\sum_{\{R_{\mu\nu}\in{Y_{n_{\mu\nu}}^{(N)}}\}}
e^{-S(\{R_{\mu\nu}\})}~
Tr_{n_{+}}[{\bf D}(~A_{n_{+}}(\{R_{\mu\nu}\})~)]~,
\label{5.1bb}
\ee
of certain master-elements $A_{n_{+}}(\{R_{\mu\nu}\})=
\sum_{\sigma\in{S(n_{+})}} a(\sigma|\{R_{\mu\nu}\})~\sigma$. Being defined by
eqs. (\ref{5.17e}),(\ref{5.17f}), they take values
in the {\it tensor} representation of the $S(n_{+})$ algebra (with
$n_{+}$ being given by eq. (\ref{2.1bx})). The latter is
deduced by linearity from the canonical representation \cite{Gr-in-phys} for
$S(n)$-group elements $\sigma$
\be
{\bf D}(\sigma)_{\{j^{\oplus n}\}}^{\{i^{\oplus n}\}}=
\delta_{j_{1}}^{i_{\sigma(1)}}
\delta_{j_{2}}^{i_{\sigma(2)}}...\delta_{j_{n}}^{i_{\sigma(n)}}~~~~;~~~~
\hat{\sigma}:~k\rightarrow{\sigma(k)}~,~k=1,...,n,
\label{3.7}
\ee
where $\delta_{j}^{i}$ denotes the '{\it N}-dimensional'
Kronecker delta function.

To relate (\ref{5.1bb}) with the stringy pattern like (\ref{0.5}), in Section
4 we represent this equation in the form suitable for the algebraic
definition of the topological data. First of all, one observes
(see Appendix B)
that the $Tr_{n_{+}}$-trace in eq. (\ref{5.1bb}) is simply related to 
the associated character of the {\it regular} $S(n_{+})$-representation.
As a result, $B(\{n_{\mu\nu}\})$ can be rewritten as the delta-function on
the $S(n_{+})$-algebra (that selects the contribution of the weight of the
$S(n_{+})$ unity-permutation $\hat{1}_{[n_{+}]}$)
\be
B(\{n_{\mu\nu}\})=Tr_{n_{+}}[{\bf D}(\tilde{A}_{n_{+}})]=
\delta_{n_{+}}(~\Lambda^{(1)}_{n_{+}}~\tilde{A}_{n_{+}}~)
\label{5.1dd}
\ee
that already played the important role in the $D=2$ analysis \cite{Gr&Tayl}.
Next, employing the Schur-Weyl duality, {\it both} the operator
$\Lambda^{(1)}_{n_{+}}$ (defined by the {\it U(N)} variant of eq. (\ref{3.8}),
see Section 3) {\it and}
\be
\tilde{A}_{n_{+}}=\sum_{\{R_{\mu\nu}\in{Y_{n_{\mu\nu}}^{(N)}}\}}
e^{-S(\{R_{\mu\nu}\})}~A_{n_{+}}(\{R_{\mu\nu}\})
\label{5.1cc}
\ee
are reformulated entirely in terms of the symmetric group elements (i.e. no
{\it SU(N)} irreps $R_{\phi}$ are left). The resulting expression (\ref{7.7})
is suitable to specify the mappings (\ref{0.1}). The
$D\geq{3}$ nature of (\ref{0.1}) is reflected by the {\it outer}-product
structure of $\tilde{A}_{n_{+}}$ which is represented as certain
combination of the
$S(n_{\phi})$-blocks embedded to act in the common
{\it enveloping} space of the $S(n_{+})=\cup_{\phi} S(n_{\phi})$ algebra
(with $\phi\in{\{\mu\nu\},\{\rho\}}$). The technique, dealing with such
compositions, is naturally inherited from the nonabelian duality
transformation \cite{Dub2}.

The remaining material is organized as following.
In Section 5, we rederive by our methods the Baez-Taylor 
reformulation \cite{Baez&Tayl} of the Gross-Taylor $2d$ pattern (\ref{0.5})
for the Heat-Kernal model (\ref{4.1b}). We also briefly sketch how our method
generalizes for any admissible model (\ref{4.1}). Being more compact than the
Gross-Taylor one, the $D=2$ representation of the Baez-Taylor type has the
structure where the full stringy pattern is not entirely manifest. To
circumvent this problem we formulate a simple prescription how to transform
the latter into the former.

The $D\geq{3}$ generalization of the $2d$ Baez-Taylor representation
\cite{Baez&Tayl} is derived in Section 6 for the 
Eguchi-Kawai models (\ref{2.1}) which fully justifies the announced pattern
(\ref{0.1}) of the $D\geq{3}$ mappings. (In particular, we make certain
conjecture concerning the concise topological reinterpretation, generalizing
the $D=2$ one \cite{Moore}, of the $D\geq{3}$ sums like (\ref{0.5}) in the
formal topological limit when $\Gamma(..)\rightarrow{0}$.) As well as in the
$D=2$ case, to relate the obtained $D\geq{3}$ representation with the
manifest stringy pattern (\ref{0.5c}), a natural extension of the $D=2$
prescription (formulated in Section 5) is suggested. In Section 7,
we discuss the structure of the asserted {\it continuous} group of the
area-preserving homeomorphisms (the generic lattice Gauge String weights
$w[\tilde{M}]$ are endowed with) and make a brief comparison with the earlier
large {\it N} variants \cite{Kaz&Zub,Kost1} of the Wilson's {\it SC} expansion
\cite{Wils} devoid of the latter invariance. In Section 8, we discuss the
major
qualitative features of the Gauge String accentuating the similarities and
differences between the continuous flux-theory and the conventional paradigm
of the $D\geq{3}$ 'fundamental' strings. In the last section, we put forward
a speculative proposal for the mechanism of confinement in the standard
weakly-coupled continuous gauge theory (\ref{0.1ew}) at large {\it N}. Also a
preliminary
contact with the two existing stringy proposals \cite{Witt2,Polyak2} is made.
Finally, the Appendices contain technical pieces of some derivations used in
the main text.

\section{The Dual Representation of $\tilde{X}_{D}$.}
\setcounter{equation}{0}

To derive (\ref{5.1bb}), we apply the nonabelian duality transformation to
the partition function of the {\it SU(N)} TEK model
(\ref{2.1})/(\ref{4.1}) following the general algorithm formulated
in \cite{Dub2} for a generic {\it SU(N)} $D$-matrix system. To begin with,
the $SU(N)$ character is to be represented in 
the form (see e.g. \cite{Moore,Dub2}) reminiscent of the one of eq.
(\ref{5.1bb}). Let $C_{R}$ denote the canonical Young idempotent
\cite{Gr-in-phys} proportional, $P_{R}=d_{R}C_{R}$, to the Young projector
\be
P_{R}={\frac{d_{R}}{n!}}\sum_{\sigma\in S(n)}\chi_{R}(\sigma)~\sigma
~~~~~,~~~~~
R\in{Y_{n}}~,
\label{3.9}
\ee
where $\chi_{R}(\sigma),~d_{R}$ are the character and the dimension associated
to the $S(n)$ irrep $R$ (while $[P_{R},\sigma]=0,~\forall{\sigma\in S(n)}$).
Then, $\chi_{R}(U)$ assumes (akin to (\ref{5.1bb})) the form of the trace:
$\chi_{R}(U)=Tr_{n} [{\bf D}(C_{R})U^{\oplus n}],~R\in{Y^{(N)}_{n}}$,
where
\be
Tr_{n} [{\bf D}(\sigma)U^{\oplus n}]=\sum_{i_{1}i_{2}..i_{n}=1}^{N}
U_{i_{1}}^{i_{\sigma(1)}}U_{i_{2}}^{i_{\sigma(2)}}...
U_{i_{n}}^{i_{\sigma(n)}}~,
\label{3.11}
\ee
and the tensor ${\bf D}(\sigma)$ is defined in
eq. (\ref{3.7}).

Altogether, it implies that the master-integral (\ref{5.1bc}) can be
rewritten in the synthetic form
\be
B(\{R_{\mu\nu}\})=\int
Tr_{4n_{+}} [{\bf D}(\Xi_{4n_{+}}(\{R_{\mu\nu}\}))~
{\bf D}(\{U_{\rho}\otimes U_{\rho}^{+}\})]
\prod_{\tilde{\rho}=1}^{D}dU_{\tilde{\rho}}
\label{5.1}
\ee
where the $S(4n_{+})$-algebra valued tensor
${\bf D}(\Xi_{4n_{+}}(\{R_{\mu\nu}\}))$ (to be defined by eq. (\ref{5.4})
below) multiplies the complementary tensor given by the ordered direct product
of the $4n_{+}$ elementary $N\times N$ matrices $(U_{\rho})^{i}_{j},~
(U_{\rho}^{+})^{k}_{l}$:
\be
{\bf D}(\{U_{\rho}\otimes U_{\rho}^{+}\})\equiv{
\bigotimes_{\rho=1}^{D}\left(~(U_{\rho})^{\oplus n_{\rho}}
\otimes (U_{\rho}^{+})^{\oplus n_{\rho}}\right)}~~~~,~~~~
2n_{+}=\sum_{\rho=1}^{D} n_{\rho}~.
\label{5.1v}
\ee
Given $\Xi_{4n_{+}}(\{R_{\mu\nu}\})$, we first derive the intermediate
$S(4n_{+})$ representation 
\be
B(\{R_{\mu\nu}\})=
Tr_{4n_{+}}[~{\bf D}(~J_{4n_{+}}(\{R_{\mu\nu}\})~)~]
\label{5.1bd}
\ee
which in Section 3.3 will be transformed into the final $S(n_{+})$ form
of eq. (\ref{5.1bb}). To determine the operator $J_{4n_{+}}\in{S(4n_{+})}$, in
(\ref{5.1}) one is to substitute the dual form \cite{Dub2} of the {\it SU(N)}
measure, i.e. to represent the result of the $D$ different
$U_{\rho}$-integrations as an $S(4n_{+})$-tensor akin to
${\bf D}(\Xi_{4n_{+}}(\{R_{\mu\nu}\}))$.

Before we explain the latter procedure, let us make the pattern of
the element $\Xi_{4n_{+}}(\{R_{\mu\nu}\})=\sum_{\sigma\in{S(4n_{+})}}
\xi(\sigma|\{R_{\mu\nu}\})~\sigma$ more transparent relating the explicit
form of the $S(4n_{+})$-group tensor ${\bf D}(\sigma)$ with the one of
(\ref{5.1v}). To this aim, we introduce first a particular
$S(4n_{+})$ permutation $\alpha_{\{n_{\rho}\}}$ via the mapping
$m\rightarrow{\alpha(m)},~m=1,...,4n_{+}$. Then, generalizing the
tensor representation (\ref{3.7}), the tensor
${\bf D}(\alpha_{\{n_{\rho}\}})$ stands for
\be
  \delta^{i_{\alpha(1)}}_{j_{1}}...
  \delta^{i_{\alpha(n_{1})}}_{j_{n_{1}}}
  \delta^{k_{\alpha(n_{1}+1)}}_{l_{n_{1}+1}}...
  \delta^{k_{\alpha(2n_{1})}}_{l_{2n_{1}}} ...
  \delta^{k_{\alpha(4n_{+}-2n_{D})}}_{l_{4n_{+}-2n_{D}}}...
  \delta^{k_{\alpha(4n_{+}-n_{D})}}_{l_{4n_{+}-n_{D}}}...
  \delta^{k_{\alpha(4n_{+})}}_{l_{4n_{+}}}~.
\label{3.18}
\ee
where, to each individual $U_{\rho}$-,$U_{\rho}^{+}$-factor in the product
(\ref{5.1v}),
we associate one copy of the Kronecker delta-function. In this way, both
${\bf D}(\alpha_{\{n_{\rho}\}})$ and the block (\ref{5.1v}) are defined to act
on one and the same $S(4n_{+})$ space (to be manifestly constructed, see eqs.
(\ref{5.13}),(\ref{5.13b}) below).
Complementary, the pattern of the trace (\ref{5.1bd}) of
$J_{4n_{+}}(\{R_{\mu\nu}\})$ (defined through the structures like
(\ref{3.18})) naturally generalizes the above convention (\ref{3.11}) to
the case of (\ref{5.1v}).

\subsection{The Dual form of the $SU(N)$ measure.}

On a given base-lattice, a generic multilink integral (like in (\ref{2.1}))
evidently can be expressed \cite{Dr&Zub} in terms of the 1-link integrals
$M^{G}(n,m)_{j_{1}...l_{m}}^{p_{1}...q_{m}}$: 
\be
\int  (U)^{p_{1}}_{j_{1}}...(U)^{p_{n}}_{j_{n}}~
(U^{+})^{q_{1}}_{l_{1}}...  (U^{+})^{q_{m}}_{l_{m}}{dU}
\equiv{\int {\bf D}(U)_{\{j^{\oplus n}\}}^{\{p^{\oplus n}\}}~
{\bf D}(U^{+})_{\{l^{\oplus m}\}}^{\{q^{\oplus m}\}}}{dU}
\label{3.5}
\ee
composed of the $N\times N$ matrices
$(U)^{p_{k}}_{j_{k}},~(U^{+})^{q_{k}}_{l_{k}}$ in the (anti)fundamental
representation of the considered Lie group $G$.
As we will show in a moment, the $SU(N)$ TEK partition function (PF)
(\ref{2.1}) is invariant under the substitution of the $SU(N)$ link-variables
by the $U(N)=[SU(N)\otimes U(1)]/{\bf Z_{N}}$ ones.
In the {\it U(N)} case, where the dual \cite{Dub2} form of
$M^{U(N)}(n,m)$ in terms of the $S(n)$-valued tensors (\ref{3.7}) reads 
\be
M^{U(N)}(n,m)_{j_{1}...l_{m}}^{p_{1}...q_{m}}=\delta[n,m]
\sum_{\sigma\in{S(n)}}{{\bf D}(\sigma^{-1} \Lambda^{(-1)}_{n})}
^{\{q^{\oplus n}\}}_{\{j^{\oplus n}\}}~
{\bf D}(\sigma)^{\{p^{\oplus n}\}}_{\{l^{\oplus n}\}}~.
\label{3.6}
\ee
that renders manifest the previously known interrelation \cite{Dr&Zub,Gr&Tayl}
between (\ref{3.5}) and the symmetric group's structures.
The operator $\Lambda^{(-1)}_{n}\in{S(n)}$ belongs to the ({\it U(N)} option
of the) family \cite{Dub2}
\be
\Lambda^{(m)}_{n}=
\sum_{R\in{Y_n^{(N)}}} d_{R}~(n!~dimR/d_{R})^{m}~C_{R}~~~,~~~m\in{\bf Z}~,
\label{3.8}
\ee
where $C_{R}=P_{R}/d_{R}$ is defined by eq. (\ref{3.9}). In eq.
(\ref{3.8}), $d_{R}$ and $dimR$ are respectively the dimension of
the $S(n)$-irrep and chiral $U(N)$-irrep both described by the same 
Young tableau $Y_n^{(N)}$ containing not more than {\it N} rows. (In eqs.
(\ref{7.5}),(\ref{7.7}) below, we will consider the {\it SU(N)} variant of
$\Lambda^{(m)}_{n}$ where the sum is traded for the one
over the {\it SU(N)} irreps.)

As for the possibility to substitute the $SU(N)$ TEK link-variables
by the $U(N)=[SU(N)\otimes U(1)]/{\bf Z_{N}}$ ones, 
the pattern of eq. (\ref{2.1}) ensures that the TEK
action is invariant under the $D$ copies of the {\it extended} transformations
$[U(1)]^{\oplus D}:~U_{\rho}\rightarrow{t_{\rho}U_{\rho}}$, where
$t_{\rho}\in{U(1)}$ rather than taking value in the center-subgroup
${\bf Z_{N}}$ of $SU(N)$. As a result, the nondiagonal moments
$M^{SU(N)}(n,m)$, $n\neq{m}$, do $not$ contribute into the TEK PF
$\tilde{X}_{D}$. As for the remaining
diagonal integrals $M^{SU(N)}(n,n)$, the latter extended invariance justifies
\cite{Dub2} the required substitution: $M^{SU(N)}(n,n)=M^{U(N)}(n,n),~
\forall{n}\in{{\bf Z}_{\geq{0}}}$. Moreover, in the context of the large
{\it N} {\it SC} expansion, in eq. (\ref{3.8}) one can
insert the {\it SU(N)} variant of $\Lambda^{(-1)}_{n}$. (The difference can be
traced back to the contributions of the {\it SU(N)} string-junctions which are
supposed to be irrelevant to all orders in {\it 1/N}.)

Actually, the derivation of (\ref{5.1bd}) calls for the alternative
$S(2n)$ reformulation \cite{Dub2} of the $S(n)\otimes S(n)$ formula
(\ref{3.6}) that will require the explicit form of an $S(2n)$-basis. For this
purpose, we first recall that
each individual matrix ${U}^{+}_{\rho}$ or ${U}_{\rho}$ can be viewed
\cite{Gr-in-phys} as the operator acting on the associated elementary
{\it N}-dimensional
subspace $|i_{\pm}(\rho)>$ according to the pattern:
$\hat{U}|i_{-}>=\sum_{j_{-}=1}^{N}U_{i_{-}}^{j_{-}}|j_{-}>$ 
and similarly for $\hat{U}^{+}|i_{+}>$.
Complementary, for a given $\sigma\in{S(n)}$, the operator
(\ref{3.7}) acts as the corresponding permutation of the elementary subspaces $|i_{k}>$
\be
\hat{\sigma}|i_{1}>|i_{2}>...|i_{n}>=
|i_{\sigma^{-1}(1)}>|i_{\sigma^{-1}(2)}>
... |i_{\sigma^{-1}(n)}>\equiv{
{\bf D}(\sigma)_{\{i^{\oplus n}\}}^{\{j^{\oplus n}\}}|j>^{\oplus n}}
\label{A.3}
\ee
where in the r.h.s. the summation $\sum_{\{j_{k}\}}$ is implied.
As a given $S(4n_{+})$-basis is constructed as the outer product of the
$4n_{+}$ building blocks $|i_{\pm}(\rho)>$ ({\it ordered} according to a
particular prescription), the elementary subspaces $|i_{k}>$ of eq. (\ref{A.3})
are represented by $|i_{\pm}(\rho)>$.

Returning to the $S(2n)$-reformulation
of eq. (\ref{3.6}), in the basis $|I_{2n}>=|I^{(+)}_{n}>\otimes |I^{(-)}_{n}>$
(with $|I^{(\pm)}_{n}>=|i_{\pm}>^{\oplus n}$) it reads
\be
\int dU {\bf D}(U)^{j_{1}...j_{n}}_{i_{1}...i_{n}}
{\bf D}(U^{+})^{j_{n+1}...j_{2n}}_{i_{n+1}...i_{2n}}=
{\bf D}(\Phi_{2n}\Gamma(2n)(\Lambda^{(-1)}_{n}\otimes{\hat{1}_{[n]}}))
^{\{j^{\oplus 2n}\}}_{\{i^{\oplus 2n}\}},
\label{5.11}
\ee
where $\hat{1}_{[n]}$ denotes
the 'unity'-permutation of the $S(n)$ group, $\Lambda^{(-1)}_{n}\in{S(n)}$
is defined by eq. (\ref{3.8}), while
$\Gamma(2n)=\sum_{\sigma\in{S(n)}}(\sigma^{-1}\otimes \sigma)$, i.e.
\be
{\bf D}(\Gamma(2n))^{\{j^{\oplus 2n}\}}_{\{i^{\oplus 2n}\}}=
\sum_{\sigma\in{S(n)}}
{{\bf D}(\sigma^{-1})}
^{j_{1}...j_{n}}_{i_{1}...i_{n}}\otimes
{{\bf D}(\sigma)^{j_{n+1}...j_{2n}}_{i_{n+1}...i_{2n}}}~\in{~S(2n)}~.
\label{5.12}
\ee
To restore the $\rho$-labels, $n\rightarrow{n_{\rho}}$, observe first that
the ordering of the $\{U_{\rho}\}$-factors in eq. (\ref{5.1v}) is
associated to the following basis
\be
|\tilde{I}_{4n(+)}>=\bigotimes_{\rho=1}^{D}|I_{2n(\rho)}>~~~;~~~
|I_{2n(\rho)}>=|I^{(+)}_{n(\rho)}>\otimes |I^{(-)}_{n(\rho)}>,
\label{5.13}
\ee
\be
|I^{(\pm)}_{n(\rho)}>=\bigotimes_{\nu\neq{\rho}}^{D-1}
|I^{(\pm)}_{n(\rho\nu)}>
~;~
|I^{(\pm)}_{n(\rho\nu)}>=|i_{\pm}(\rho)>^{\oplus n(\rho\nu)}
|i_{\pm}(\nu)>^{\oplus n(\rho\nu)},
\label{5.13b}
\ee
where $2n_{+}=\sum_{\rho=1}^{D} n_{\rho}$,
and $|I^{(\pm)}_{n}>=|i_{\pm}>^{\oplus n}$ (used in eq. (\ref{5.11})) matches
with $|I^{(\pm)}_{n(\rho)}>$. Therefore, for a given link $\rho$, the left and the
right $S(n_{\rho})$-subblocks of $\Gamma(2n_{\rho})$ in eq. (\ref{5.12}) act
respectively on $|I^{(+)}_{n(\rho)}>$ and on $|I^{(-)}_{n(\rho)}>$.
The same convention is used for the $S(n_{\rho})$-subblocks
in the direct product
$(\Lambda^{(-1)}_{n_{\rho}}\otimes{\hat{1}_{[n_{\rho}]}})$
entering eq. (\ref{5.11}).

The remaining $S(2n_{\rho})$-operator $\Phi_{2n_{\rho}}$,
being considered in the alternatively ordered basis $|\tilde{I}_{2n(\rho)}>$
for each $|I_{2n(\rho)}>$-subsector, 
\be
|I_{2n(\rho)}>\rightarrow{|\tilde{I}_{2n(\rho)}>}
=(|i_{+}(\rho)>\otimes |i_{-}(\rho)>)^{\oplus n_{\rho}}
\label{5.14}
\ee
(with $|i_{\pm}(\rho)>^{\oplus {n}(\rho)}=
\otimes_{\nu\neq{\rho}}^{D-1}|i_{\pm}(\rho)>^{\oplus n(\rho\nu)}$), 
takes the simple form of the outer product of the
2-cycle permutations $c_{2}\in{C(2)}$
\be
\Phi_{2n_{\rho}}={(c_{2})}^{\oplus n_{\rho}}\in{S(2n_{\rho})}~~~;~~~
c_{2}:\{12\}\rightarrow{\{21\}}~,
\label{5.15}
\ee
where each $c_{2}\in{S(2)}$ acts on the 'elementary'
sector $|i_{+}(\rho)>\otimes |i_{-}(\rho)>$.
It completes the embedding of the $S(2n_{\rho})$ operators (\ref{5.11}),
representing the individual 1-link integrals, to act in the common
'enveloping' $S(4n_{+})$-space. (It is noteworthy \cite{Dub2}
that the three $S(2n_{\rho})$ subblocks of the inner-product in the r.h.s. of
eq. (\ref{5.11}) commute with each other.)

\subsection{The Dual form of the TEK action.}

Let us now turn to the derivation of the master-element
$\Xi_{4n_{+}}(\{R_{\mu\nu}\})$ entering the synthetic representation
(\ref{5.1}) of the master-integral (\ref{5.1bc}). For this purpose, we first
specify an alternative, more suitable  $S(4n_{+})$ basis
\be
|I_{4n(+)}>=\bigotimes_{\mu\nu=1}^{D(D-1)/2}|I_{4n(\mu\nu)}>~,
\label{5.6}
\ee
\be
|I_{4n(\mu\nu)}>
=(|i_{+}(\mu)>|i_{+}(\nu)>|i_{-}(\mu)>|i_{-}(\nu)>)^{\oplus n_{\mu\nu}}
\label{5.6b}
\ee
where in eq. (\ref{5.6b}) the product of the four elementary blocks
$|i_{\pm}(\rho)>$ is associated to the elementary $\mu\nu$-holonomy
$U_{\mu\nu}$ entering eq. (\ref{5.1bc}).
As it is derived in Appendix A, in this basis one obtains
\be
\Xi_{4n_{+}}(\{R_{\mu\nu}\})=
\bigotimes_{\mu\nu=1}^{D(D-1)/2}
 P_{4n_{\mu\nu}}(R_{\mu\nu})\cdot \Psi_{4n_{\mu\nu}} ~~~;~~~
n_{+}=\sum_{\{\mu\nu\}} n_{\mu\nu}~,
\label{5.4}
\ee
where each of the operators
$P_{4n_{\mu\nu}}(R_{\mu\nu}),~\Psi_{4n_{\mu\nu}}\in{S(4n_{\mu\nu})}$ 
is supposed to act on the corresponding
$|I_{4n(\mu\nu)}>$ subspace of $|I_{4n(+)}>$. Given (\ref{5.6b}),
$\Psi_{4n_{\mu\nu}}$ assumes the simple form of the outer product
\be
\Psi_{4n_{\mu\nu}}=(c_{4})^{\oplus n_{\mu\nu}}\in{S(4n_{\mu\nu})}~~~~~;~~~~~
c_{4}:~\{1234\}\rightarrow{\{4123\}}~,
\label{5.7}
\ee
with each individual 4-cycle permutation $c_{4}\in{C(4)}$ acting on the
elementary plaquette subspace
\be
|i_{+}(\mu)>|i_{+}(\nu)>|i_{-}(\mu)>|i_{-}(\nu)>
\label{5.7b}
\ee
ordered in accordance with the original pattern (\ref{5.1bc}) of the
plaquette-labels of $U_{\mu\nu}$.

As for $P_{4n_{\mu\nu}}(R_{\mu\nu})$, making use of the alternative basis
$|I_{4n(\mu\nu)}>\rightarrow{|\tilde{I}_{4n(\mu\nu)}>}$ of the
$S(4n_{\mu\nu})$-subspace in (\ref{5.6b}):
\be
|\tilde{I}_{4n(\mu\nu)}>=|i_{+}(\mu)>^{\oplus n_{\mu\nu}}
|i_{+}(\nu)>^{\oplus n_{\mu\nu}}
|i_{-}(\mu)>^{\oplus n_{\mu\nu}}|i_{-}(\nu)>^{\oplus n_{\mu\nu}}~,
\label{5.9}
\ee
we employ (see Appendix A) the following representation
\be
P_{4n_{\mu\nu}}(R_{\mu\nu})=
\hat{1}_{[n_{\mu\nu}]}\otimes \hat{1}_{[n_{\mu\nu}]}
\otimes (C_{R_{\mu\nu}}\sqrt{d_{R_{\mu\nu}}})\otimes
(C_{R_{\mu\nu}}\sqrt{d_{R_{\mu\nu}}})~,
\label{5.10}
\ee
where $\hat{1}_{[n_{\mu\nu}]}$ denotes the
$S(n_{\mu\nu})$-unity. More explicitly, the four (ordered)
$S(n_{\mu\nu})$-factors in eq. (\ref{5.10}) are postulated to act on
the corresponding four (ordered) $n_{\mu\nu}$-dimensional subspaces
(\ref{5.9}) of $|\tilde{I}_{4n_{\mu\nu}}>$. Altogether, we have
formulated the required (for the derivation of eq. (\ref{5.1bd})) $embedding$
of the $S(4n_{\mu\nu})$ operators $\Psi_{4n_{\mu\nu}},~
P_{4n_{\mu\nu}}(R_{\mu\nu})$ into
the {\it enveloping} $S(4n_{+})$-space.

\subsection{$B(\{R_{\mu\nu}\})$ as the
$Tr_{n_{+}}$-character.}

Combining together eqs. (\ref{5.4}) and (\ref{5.11}) in compliance
with the pattern of eq. (\ref{5.1}), we arrive at the
{\it dual} representation of the master-integral (\ref{5.1bc}) as
the $S(4n_{+})$ {\it character} (\ref{5.1bd}) with the 
master-element
\be
J_{4n_{+}}(\{R_{\mu\nu}\})=
\left(\otimes_{\mu\nu=1}^{D(D-1)/2}~\Psi_{4n_{\mu\nu}}\right)\cdot
\left(\otimes_{\rho=1}^{D} \Delta_{2n_{\rho}}(\{R_{\rho\nu}\})\right),
\label{5.17}
\ee
\be
\Delta_{2n_{\rho}}(\{R_{\nu\rho}\})=\Phi_{2n_{\rho}}
\cdot \Gamma(2n_{\rho})\cdot K_{2n_{\rho}}(\{R_{\nu\rho}\})~~
\in{~~S(2n_{\rho})}~.
\label{5.18}
\ee
For later convenience, we have introduced the $S(n_{\rho})\bigotimes
S(n_{\rho})$ combination that in the $|I_{2n(\rho)}>$ basis (\ref{5.13})
reads
\be
K_{2n_{\rho}}(\{R_{\nu\rho}\})=\Lambda^{(-1)}_{n_{\rho}}
\bigotimes{\left(\otimes_{\nu\neq{\rho}}^{D-1}
{C_{R_{\rho\nu}}\sqrt{d_{R_{\rho\nu}}}} \right)}~,
\label{5.19}
\ee
where $\Lambda^{(-1)}_{n_{\rho}}$ acts onto 
$|I^{(+)}_{n(\rho)}>$, while each of the $C_{R_{\rho\nu}}$ factors acts
on the $|I^{(-)}_{n(\rho\nu)}>$-subspace of $|I^{(-)}_{n(\rho)}>$.

Next, let us complete the duality transformation trading the intermediate
$S(4n_{+})$ representation (\ref{5.1bd}) for its final $S(n_{+})$ pattern
(\ref{5.1bb}). To this aim, it is convenient to start with the alternative 
following alternative form of the master-element (\ref{5.17}).
As it is demonstrated in Appendix A, the dual
representation (\ref{5.1bd}) does {\it not} alter when
in the element (\ref{5.17})/(\ref{5.18}) one makes the substitution 
\be
\otimes_{\rho=1}^{D}\Phi_{2n_{\rho}}\rightarrow{\otimes_{\rho=1}^{D}
(\Phi_{2n(\rho)})^{2} }
\cong{\otimes_{\rho=1}^{D} \hat{1}_{[2n_{\rho}]}}=\hat{1}_{[4n_{+}]}~.
\label{5.21}
\ee
so that (inside the $Tr_{4n_{+}}$-character) in eq. (\ref{5.18})
all the operators $\Phi_{2n(\rho)}$ can be omitted.

\subsubsection{The $D=2$ case.}

Next, it is appropriate to proceed with the simplest $D=2$ case of
(\ref{5.1bd}) corresponding to the well studied continuous {\it SU(N)} gauge
theory on a 2-torus. It will provide not only with a cross-check of our
$D\geq{2}$ formalism but also with the motivation for the announced
reduction, $S(4n_{+})\longrightarrow{S(n_{+})}$, of the enveloping space.
In $D=2$, the this reduction is encoded in the basic property (following from
eqs. (\ref{BB.1}),(\ref{C.3}) in Appendix A) of the
$\Psi_{4n}$ operator (\ref{5.7})
\be
Tr_{n}[U_{\mu\nu}^{\oplus n}]=Tr_{4n}[{\bf D}(\Psi_{4n})
\tilde{U}_{\mu\nu}^{\oplus n}]~~~;~~~
\tilde{U}_{\mu\nu}=
U_{\mu}\otimes U^{+}_{\mu}\otimes U_{\nu}\otimes U^{+}_{\nu}~,
\label{5.21bb}
\ee
while $U_{\mu\nu}=U_{\mu}\cdot U_{\nu}\cdot U^{+}_{\mu}\cdot U^{+}_{\nu}$.
Making the substitution $U_{\rho}\rightarrow{\sigma^{(+)}_{\rho}},~
U^{+}_{\rho}\rightarrow{\sigma^{(-)}_{\rho}}$, one obtains
\be
Tr_{4n}[{\bf D}(\left(\bigotimes_{\rho=1}^{2}
(\sigma^{(+)}_{\rho}\otimes \sigma^{(-)}_{\rho}) \right)\cdot \Psi_{4n})]=
Tr_{n}[{\bf D}(\prod_{\rho=1}^{D}\sigma^{(+)}_{\rho}\cdot
\prod_{\mu=1}^{D} \sigma^{(-)}_{\mu})],
\label{5.21b}
\ee
where in the l.h.s. the operators
$\sigma^{(\pm)}_{\rho}\in{S(n)}$
(combined into the {\it outer} product) act on the associated
$|I^{(\pm)}_{n}>$ subspaces of the $S(4n)$ basis (\ref{5.13}). As 
for the ordering inside the two {\it inner} $\rho$-products in the r.h.s. of
eq. (\ref{5.21b}), in both products it complies with the ordering of the
$|i_{+}(\rho)>$ (or, equally, $|i_{-}(\rho)>$) elementary blocks in eq.
(\ref{5.7b}).

Let us apply the identity (\ref{5.21b}) to the $D=2$ option of (\ref{5.1bd}),
(\ref{5.17}). Combining eq. (\ref{5.21b}) with 
the orthonormality $P_{R_{1}}P_{R_{2}}=\delta_{R_{1},R_{2}} P_{R_{1}}$
of $P_{R}=d_{R}C_{R}$ and taking into account the standard relation
$Tr_{n}[C_{R}\sigma]=dimR\chi_{R}(\sigma)/d_{R}$ (between the projected
$Tr_{n}$-trace and the canonical $S(n)$ character $\chi_{R}$, see e.g.
\cite{Dub2}), one easily obtains 
\be
B(R)=\frac{1}{dimR}~\{~\left(\frac{d_{R}}{n!}\right)^{2}
\sum_{\{\sigma_{\rho}\in{S(n)}\}}
\frac{\chi_{R}([\sigma_{1},\sigma_{2}])}{d_{R}}~\}=\frac{1}{dimR}~,
\label{5.21e}
\ee
where $R\in{Y_{n}^{(N)}}$, and $[\sigma_{1},\sigma_{2}]$
conventionally stands for $(\sigma_{1}\sigma_{2}
\sigma^{-1}_{1}\sigma^{-1}_{2})$. We have also used that the block in the
curly brakets of (\ref{5.21e}) is
equal to unity according to the identity derived in \cite{Gr&Tayl}. 
Together with eq. (\ref{2.1b}), the expression (\ref{5.21e})
for $B(R)$ precisely matches with the (genus one) result of \cite{Rus,Witt91}
derived by the combinatorial method of \cite{Migd75}.

\subsubsection{The $D\geq{3}$ case.}

Returning to the generic $D\geq{3}$ $Tr_{4n_{+}}$-character (\ref{5.1bc}) of
the master-element (\ref{5.17}), the $S(4n_{+})\rightarrow{S(n_{+})}$
reduction of the enveloping space is performed with the help
of the following generalization of the $D=2$ identity (\ref{5.21b})
$$
Tr_{4n_{+}}[~
{\bf D}(\left(\bigotimes_{\mu\nu=1}^{D(D-1)/2}\Psi_{4n_{\mu\nu}}\right)\cdot
\left(\bigotimes_{\rho=1}^{D}~(\sigma^{(+)}_{\rho}\otimes \sigma^{(-)}_{\rho})
\right))~]=
$$
\be
=Tr_{n_{+}}[~{\bf D}(
\left(\prod_{\rho=1}^{D}
(\sigma^{(+)}_{\rho} \otimes \hat{1}_{[\frac{n_{+}}{n_{\rho}}]})\right)\cdot
\left(\prod_{\rho=1}^{D}
(\sigma^{(-)}_{\rho}\otimes \hat{1}_{[\frac{n_{+}}{n_{\rho}}]})\right))~].
\label{5.17b}
\ee
Let us simply explain the meaning of the above pattern, while for more details 
see Appendix A. In the l.h.s. of (\ref{5.17b}), the {\it outer}
$\mu\nu$-product is defined in the same way as in eq. (\ref{5.17}), and the
operators $\sigma^{(\pm)}_{\rho}\in{S(n_{\rho})}$ (composed into the
{\it outer} $\rho$-product) act on the associated $|I^{(\pm)}_{n(\rho)}>$
subspace of the $S(4n_{+})$ basis (\ref{5.13}). As for the r.h.s. of
(\ref{5.17b}), we first construct the following $S(n_{+})$ basis. To begin
with , one is to introduce the $N$-dimensional spaces
$|i(\mu\nu)>,~i=1,...,N,$ parametrized by the
plaquette label $\mu\nu=1,...,D(D-1)/2$. Then, the $S(n_{+})$ operators
(recall that $n_{+}=\sum_{\{\mu\nu\}} n_{\mu\nu}$) can be viewed as
acting on
\be
|I_{n(+)}>=\bigotimes_{\mu\nu=1}^{D(D-1)/2}|I_{n(\mu\nu)}>~~~,~~~~
|I_{n(\mu\nu)}>=(|i(\mu\nu)>)^{\oplus n_{\mu\nu}}~,
\label{5.6c}
\ee
according to the same rule (\ref{A.3}) that has been already used for the
$S(4n_{+})$ operators. Given this convention, each operator
$\sigma^{(\pm)}_{\rho}$ is postulated to act on the associated $S(n_{\rho})$
subspace $|\tilde{I}_{n(\rho)}>$  of $|I_{n(+)}>$,
\be
|\tilde{I}_{n(\rho)}>=\bigotimes_{\nu\neq{\rho}}^{D-1}
~(|i(\rho\nu)>)^{\oplus n_{\rho\nu}}~~~,~~~n_{\rho}=
\sum_{\nu\neq{\rho}}^{D-1} n_{\rho\nu}~,
\label{5.6d}
\ee
where the ordering of the $|i(\rho\nu)>$ blocks matches with the one in
eq. (\ref{5.6c}). As for
$\hat{1}_{[{n_{+}}/{n_{\rho}}]}$, in eq. (\ref{5.17b}) it denotes the unity
permutation on the $S(n_{+}-n_{\rho})$ subspace of (\ref{5.6c}) complementary
to (\ref{5.6d}). In the $D=2$ case, where $n_{+}=n_{1}=n_{2}$, the general
eq. (\ref{5.17b}) readily reduces to its degenerate variant (\ref{5.21b}).

To complete the construction (\ref{5.17b}), one should specify the ordering
inside the two inner $\rho$-products of its r.h. side. We defer this task till
the end of the section and now apply the identity (\ref{5.17b}) to the
$Tr_{4n_{+}}$ character (\ref{5.1bd}) of the master-element (\ref{5.17}).
For this purpose, all what we need is the proper identification of the
composed into $J_{4n_{+}}(\{R_{\mu\nu}\})$ permutations with
$\sigma^{(\pm)}_{\rho}$. To this aim, let denote
by $\lambda_{\mu\nu}$ and $\lambda_{\rho}$ the permutations which enter the
definition (\ref{3.9}) of the relevant operators
$C_{R_{\phi}}=P_{R_{\phi}}/d_{R_{\phi}}\in{S(n_{\phi})}$ (combined into
$K_{2n_{\rho}}$ of
eq. (\ref{5.19}), with $\Lambda^{(-1)}_{n_{\rho}}\in{S(n_{\rho})}$ being given
by eq. (\ref{3.8})) assigned with the associated labels
$\phi\in{\{\mu\nu\},\{\rho\}}$.
Complementary, let $\sigma_{\rho}$ stands for the permutations
entering the definition (\ref{5.12}) of $\Gamma(2n_{\rho})$. Identifying
$(\otimes_{\mu\neq{\rho}}^{D-1}\lambda_{\mu\rho})\cdot
\sigma_{\rho}\rightarrow{\sigma^{(-)}_{\rho}},~
\lambda_{\rho}\cdot \sigma^{-1}_{\rho}\rightarrow{\sigma^{(+)}_{\rho}}$
(without numb summation over $\rho$), after some routine machinery one
derives for the master-element
\be
A_{n_{+}}(\{R_{\mu\nu}\})=\prod_{\rho=1}^{D}
\sum_{R_{n_{\rho}}\in{Y_{n_{\rho}}^{(N)}}}
\frac{d^{2}_{R_{\rho}}}{n_{\rho}!dimR_{\rho}}
\sum_{\sigma_{\rho}\in{S(n_{\rho})}} F(\{\sigma_{\rho}\};\{R_{\phi}\})~,
\label{5.17e}
\ee
\be
F=\left(\bigotimes_{\{\mu\nu\}} C_{R_{\mu\nu}}\right) \cdot
\left(\prod_{\{\rho\}}
(\sigma_{\rho} \otimes \hat{1}_{[\frac{n_{+}}{n_{\rho}}]})\right)\cdot
\left(\prod_{\{\lambda\}}
(\sigma^{-1}_{\lambda}C_{R_{\lambda}}\otimes
\hat{1}_{[\frac{n_{+}}{n_{\lambda}}]})\right),
\label{5.17f}
\ee
so that its trace (\ref{5.1bb}) determines the $B(\{R_{\mu\nu}\})$-block
(\ref{5.1bc}) of the TEK partition function (PF).

Altogether, it establishes the exact duality transformation of
the TEK  PF (\ref{2.1}) which is one of the main
results of the paper. In certain sense (modulo
the explicit presence of the $\sigma_{\rho}$-twists), it provides with the
$D\geq{3}$ generalization of the representation \cite{Rus,Witt91} for the PF
of the continuous gauge theory on a $2d$ manifold. Also, it can be compared
with the considerably simpler pattern of the PF of the judiciously constructed
solvable $D$-matrix models \cite{Dub1} where $B(\{R_{\mu\nu}\})$ depends
nontrivially only on the associated generalized Littlewood-Richardson
coefficients. (On the contrary, the pattern (\ref{5.17}) encodes the general
Klebsch-Gordan coefficiens.)

\subsubsection{The ordering inside the inner $\rho$-products .}

Finally, let us discuss the ordering inside the two inner
$\rho$-products of the r.h. side of eq. (\ref{5.17b}). As it is shown in
Appendix A, this ordering is entirely predetermined by the following 
characteristics $G(\rho)$ of a particular $\rho$-label called its
{\it cardinality}. Consider the $D(D-1)/2$ dimensional vector
${\bf M}(\{\rho\})$ defined so that its components $M_{k}(\{\rho\})$ are
in one-to-one correspondence with the $D(D-1)/2$ labels $\mu\nu$ (where
$1\leq{\mu}<\nu\leq{D}$). The latter parametrize the $\mu\nu$th
plaquette-holonomies
$U_{\mu\nu}$ entering the characters in eq. (\ref{5.1bc}). Let the $\mu\nu$th
component of ${\bf M}(\{\rho\})$ is equal to the first link-label $\mu$ of
the plaquette-label: $M_{\mu\nu}(\{\rho\})=\mu$.
Then, the cardinality (ranging from $0$ to $D-1$) is postulated to be
$G(\mu)=\sum_{k=1}^{D(D-1)/2} \delta [\mu,M_{k}(\{\rho\})]$, i.e.
the number of times the particular
$\mu$-label enters the entries of the vector ${\bf M}(\{\rho\})$. In eq.
(\ref{5.1bc}), it is always possible to arrange for a {\it nondegenerate}
cardinality assignement $\{G(\rho)\}$ when $G(\rho_{1})\neq{G(\rho_{2})}$ if
$\rho_{1}\neq{\rho_{2}}$. Then, {\it given a nondegenerate set $\{G(\rho)\}$,
the $(\sigma^{(\pm)}_{\rho} \otimes \hat{1}_{[{n_{+}}/{n_{\rho}}]})$ factors
in the eq. (\ref{5.17b}) are ordered (from the left to the right) according
to the successively decreasing $G(\rho)$-assignements of their $\rho$-labels}.

\section{ Schur-Weyl transformation of $B(\{n_{\mu\nu}\})$.}

To transform the dual representation (\ref{5.1bb})/(\ref{5.17e}) of the TEK
PF $\tilde{X}_{D}$ into the $D\geq{2}$ stringy representation like
(\ref{0.5}), one is rewrite $B(\{n_{\mu\nu}\})$ entirely in terms
of the symmetric groups' variables which are suitable for the algebraic
definition of the topological data associated to the mappings (\ref{0.1}). For
this purpose, we employ the following two useful identities (derived in
Appendices B and D) which reflect the Schur-Weyl complementarity of the Lie
and the symmetric groups. The first one trades the ubiquitous operator
(\ref{3.8}) for the product of the two elements of the associated $S(n)$
algebra
\be
\Lambda^{(m)}_{n}=P_{n}^{(N)}\cdot (N^{n}\Omega_{n})^{m}~~~~~,~~~~~~
P_{n}^{(N)}=\sum_{R\in{Y_n^{(N)}}} P_{R}~,
\label{7.1}
\ee
where the projector $(P_{n}^{(N)})^{2}=P_{n}^{(N)},~P_{n}^{(N)}=1$ if $n<N$,
(that independently appeared within the method of \cite{Baez&Tayl})
reduces the $S(n)$ $Y_{n}$-variety of irreps to the $Y_n^{(N)}$-one of either
{\it U(N)} or $SU(N)$. As for the second element 
\be
\Omega_{n}=\sum_{\sigma\in{S(n)}} (1/N)^{n-K_{[\sigma]}}~\sigma~~~~~;~~~~~
[~\Omega_{n},\rho~]=0~,~\forall{\rho\in{S(n)}},
\label{7.3}
\ee
(belonging the center of the $S(n)$-algebra), it is defined \cite{Gr&Tayl} by
the equation
\be
(dimR)^{m}=
\frac{\chi_{R}((N^{n}\Omega_{n})^{m})}{d_{R}}
\left(\frac{d_{R}}{n!}\right)^{m}~~~;~~~
n-K_{[\sigma]}=\sum_{k=1}^{n} (k-1)p_{k}~,
\label{7.4}
\ee
where $m\in{\bf Z}$, and the factor $K_{[\sigma]}=\sum_{k=1}^{n} p_{k}$ in
eq. (\ref{7.3}) denotes the total number of various $k$-cycles in the cyclic
decomposition of the conjugacy class $[\sigma]=[1^{p_1},2^{p_2},...,n^{p_n}]$,
$\sum_{k=1}^{n} kp_{k}=n$.

Generalizing (\ref{7.1}), the second key-identity  deals with the similar
sum weighted this time by the factor $e^{-\Gamma}$ (which defines a generic
model (\ref{4.1}))
\be
\sum_{R\in{Y_n^{(N)}}}
e^{-\Gamma(..,\{C_{p}(R)\})}~d_{R}
\left(\frac{n!dimR}{d_{R}}\right)^{m}C_{R}=
P_{n}^{(N)}\cdot (N^{n}\Omega_{n})^{m}\cdot Q_{n}(\Gamma).
\label{7.5}
\ee
In the Heat-Kernal case (\ref{4.1b}) where $\Gamma(..)=\lambda C_{2}(R)/2N$,
the $S(n)$-algebra valued operator $Q_{n}(\Gamma)$ reads
\be
Q_{n}(\Gamma)=exp[{-\frac{\lambda}{2}(n-\frac{n^2}{N^2})}]~
exp[{-{\frac{\lambda}{N}\hat{T}^{(n)}_{2}}}]~~~~,~~~~\hat{T}^{(n)}_{2}=
\sum_{\tau\in{T^{(n)}_{2}}} \tau~,
\label{7.6}
\ee
where $T^{(n)}_{2}\equiv{T_{2}}$ denotes the $S(n)$ conjugacy
class $[1^{n-2}2^{1}]$ of the simple transposition, and
$\lambda=\tilde{g}^2N$. In a generic admissible model
(\ref{4.1}), as it is demonstrated in Appendix E, the operator
$Q_{n}(\Gamma)$ generalizes to
\be
Q_{n}(\Gamma)=exp[\sum_{{\{p\}}}
\upsilon_{\{p\}}(\{\tilde{b}_{k}\},n,N)~\hat{T}^{(n)}_{\{p\}}]
~~~,~~~
\hat{T}^{(n)}_{\{p\}}=\sum_{\xi^{\{p\}}\in{T^{(n)}_{\{p\}}}} \xi^{\{p\}},
\label{7.6b}
\ee
where $\hat{T}^{(n)}_{\{p\}}$ denotes the sum of the
$S(n)$ permutations belonging to a particular conjugacy class
${T}^{(n)}_{\{p\}}$ labelled by
the partition ${\{p\}}$ of $n$: $\sum_{i=1}^{n} kp_{k}=n,$. As for the
weight $\upsilon_{\{p\}}(..)$, it assumes the form (which, in particular,
results in the required asymptotics (\ref{0.1ef}))
\be
\frac{\upsilon_{\{p\}}(\{\tilde{b}_{k}\},n,N)}{N^{-\sum_{k=1}^{n} (k-1)p_{k}}}
=\sum_{m=0}^{M_{\{p\}}}\sum_{l\geq{[m/2]}}
~s_{\{p\}}(\{\tilde{b}_{k}\},m,l)~N^{-2l}~n^{m},
\label{7.6c}
\ee
where $m,l\in{\bf Z_{\geq{0}}}$, and $[m/2]=m/2$ or $(m+1)/2$ depending on
whether $m$ is even or odd (while
$s_{\{p\}}(..)\sim{O(N^{0})}$). The specific
pattern of $\upsilon_{\{p\}}(..)$ implies that in eq. (\ref{4.1}) the
function $\Gamma(..)$ satisfies certain conditions (see Appendix E)
that ensure the
consistent stringy interpretation of 
(\ref{7.6c}) to be discussed in Section 6 (for eq. (\ref{7.6})) and in
Appendix E (for eqs. (\ref{7.6b})/(\ref{7.6c})).

Combining (\ref{7.1}),(\ref{7.5}) with the delta-function reformulation
(\ref{5.1dd}) of the $Tr_{n_{+}}$ character (derived in Appendix B), 
we arrive at the explicit $\otimes_{\phi} S(n_{\phi})$ representation
for the building block $B(\{n_{\mu\nu}\})$ of the TEK partition
function (\ref{2.1b}). This central, for the present discussion
of the lattice Gauge String, expression reads
\be
\delta_{n_{+}}(\Lambda_{n_{+}}
\left(\bigotimes_{\{\mu\nu\}} \frac{Q_{n_{\mu\nu}}\Lambda_{n_{\mu\nu}}}
{n_{\mu\nu}!}\right) \sum_{\{\sigma_{\tilde{\rho}}\}}
\prod_{\{\rho\}} 
(\sigma_{\rho} \otimes \hat{1}_{[\frac{n_{+}}{n_{\rho}}]})
\prod_{\{\mu\}}
(\sigma^{-1}_{\mu}\Lambda^{(-1)}_{n_{\mu}}\otimes
\hat{1}_{[\frac{n_{+}}{n_{\mu}}]})),
\label{7.7}
\ee
where, to all orders in {\it 1/N}, one can safely use the {\it SU(N)} variant
of the representation (\ref{7.1}) of  $\Lambda^{(m)}_{n_{\phi}}$.
It is noteworthy that the $[{\bf Z_{2}}]^{\oplus D(D-1)/2}$ invariance
(with respect to $R_{\mu\nu}\leftrightarrow{\bar{R}_{\mu\nu}}$) of the sums
in (\ref{2.1}) defining the plaquette-factor (\ref{4.1}) results in the
invariance of eq. (\ref{7.7}) under the simultaneous permutations $\rho,\mu
\rightarrow{\sigma(\rho),\sigma(\mu)},~\forall{\sigma}\in{S(D)}$
of the link-labels $\rho,\mu=1,...,D,$ in the two ordered inner products.

\section{The stringy form of $B(n)$ in $D=2$.}

To begin with, in the $D=2$ case (where $n_{+}=n_{12}=n_{1}=n_{2}$) all the
involved into (\ref{7.7}) $S(n_{\phi})$ operators act
in one and the same $S(n)$-space that matches with the reduced formula
(\ref{5.21b}). The resulting amplitude in the Heat-Kernal case (\ref{4.1b})
reads (with $B(0)\equiv{1}$)
\be
B(n)=\frac{e^{-\frac{\lambda}{2}(n-\frac{n^2}{N^2})}}{n!}
\sum_{\{\sigma_{\rho}\},T^{(n)}_{2}\in{S(n)}}
\delta_{n}( P_{n}^{(N)} e^{-{\frac{\lambda}{N}\hat{T}^{(n)}_{2}}} 
(N^{n}\Omega_{n})^{1-2+1} [\sigma_{1},\sigma_{2}])
\label{7.8}
\ee
that is in complete agreement with the genus-one result \cite{Baez&Tayl} of
Baez and Taylor derived by a different method. It provides with the
more compact reformulation of the Gross-Taylor stringy pattern (\ref{0.5}),
although the transformation (to be summarized by eqs.
(\ref{7.8b}),(\ref{7.8c}) below) relating the two representations
is not entirely manifest. Let us proceed recasting (\ref{7.8}) into the form
'almost' equivalent to the one of (\ref{0.5}).

To make
contact with the pattern of the $f$-mapping of eq. (\ref{0.5}), first it
is convenient to decompose
\be
P_{n}^{(N)}=
\sum_{T^{(n)}_{\{p\}}\in{S(n)}}P_{n}^{(N)}(T^{(n)}_{\{p\}})~
\hat{T}^{(n)}_{\{p\}}~,
\label{7.11xbx}
\ee
where $\hat{T}^{(n)}_{\{p\}}$ is defined in eq.
(\ref{7.6b}). Then, expanding all the exponents except
$e^{-\frac{\lambda}{2}n}=e^{-\tilde{\sigma}_{0}n}$, one is to rewrite
(\ref{7.8}) as
\be
B(n)=\sum_{i,s,t\geq{0}} \sum_{T_{\{p\}}\in{S(n)}}~
\sum_{f\in{\tilde{M}}}
\frac{N^{-2(t+s)-i}}{|C_{f}(\{p\})|}
P_{n}^{(N)}(T_{\{p\}})~K_{n}(i,s,t),
\label{7.11}
\ee
\be
K_{n}(i,s,t)=e^{-\frac{\lambda}{2}n}~
\frac{(\lambda)^{i+s+t}}{i!s!t!}~\frac{(-1)^{i}n^{s}
(n^{2}-n)^{t}}{2^{s+t}}~,
\label{7.11b}
\ee
where in the first exponent of (\ref{7.8}) one is to decompose $n^{2}=n/2+
n(n-1)/2$, and for simplicity we denote
$T_{\{p\}}\equiv{T^{(n)}_{\{p\}}}$ in
the rest of the section. As for the symmetry factor,
\be
\sum_{f\in{\tilde{M}(\{p\},n,i)}}
\frac{1}{|C_{f}(\{p\})|}=
\sum_{\{\sigma_{\rho}\in{S(n)}\}}
\frac{1}{n!}~\delta_{n}(~\hat{T}_{\{p\}}~(\hat{T}_{2})^{i}~
[\sigma_{1},\sigma_{2}]~),
\label{7.12}
\ee
it emerges when one reformulates the r.h.s. of (\ref{7.12}) as the sum over
certain maps (\ref{0.5}) (to be explicitly constructed below). The latter can be viewed as
the topological mappings (i.e. immersions {\it without} singularities)
\be
f~:~~~~~f(\tilde{M}-\{f^{-1}(q_{s})\})=M-\{q_{s}\}~.
\label{7.11bb}
\ee
of the space $\tilde{M}-\{f^{-1}(q_{s})\}$ onto the base-space torus {\it M}
with $i+1$ deleted points $\{q_{s}\}$.
These maps define \cite{Massey} the admissible (by the data in the
r.h.s. of (\ref{7.12})) branched covering spaces $\tilde{M}$ of {\it M} which
can be visualized as the Riemann surfaces
$\tilde{M}\equiv{\tilde{M}(\{p\},n,i)}$ (to be identified with
the worldsheets of the {\it YM}-flux) with $n$-sheets and
$i+1$ branch points located at $\{q_{s}\}$.

As for the sums (\ref{7.11}) over the 
nonnegative integers $i,t,s$, in addition to the number of the
'movable' simple branch-points (where two 
sheets are identified), they refer to the extra 'movable' 
singularities of the map (\ref{0.5}). Namely, one is to attach to 
$\tilde{M}(\{p\},n,i)$ the $t$ collapsed to a point microscopic tubes 
(connecting two sheets) and the $s$ collapsed to a point handles 
(glued to a single sheet). Altogether, it results in the worldsheet 
$\tilde{M}(\{p\},n,i|t,s)$. 

The important observation \cite{Gr&Tayl} is that the factor
$|C_{f}(\{p\})|$ is equal to the number of distinct automorphisms of 
the branched covering space $\tilde{M}(\{p\},n,i|t,s)$ in question. 
Complementary, in the absence of the $P_{n}^{(N)}$-twist (i.e. when
$T_{\{p\}}\rightarrow{\hat{1}}$),
the {\it 1/N} factor enters (\ref{7.11}) in the power equal to the 
$G=1$ option of the Riemann-Hurwitz formula $h=n\cdot (2G-2)+2(t+s)+i$
calculating the overall genus of the corresponding (modified) Riemann 
surface $\tilde{M}(\hat{1},n,i|t,s)$. Also, in what follows, we assume that
that the contribution of the 'movable' collapsed handles is reabsorbed
into the redefinition $\tilde{\sigma}_{0}=\lambda/2
\rightarrow{\lambda(1-1/N^{2})/2}$ of the $SU(N)$ bare string tension 
which eliminates the corresponding singularities of the map 
(\ref{0.1}). (In the case (\ref{7.6b}) of the generic model 
(\ref{4.1}), the modified tension is given by the $n=1$ restriction of 
$-\upsilon_{\{p\}}(\{\tilde{b}_{k}\},n,N)$ associated to 
$\hat{T}^{(n)}_{\{p\}}=\hat{1}_{[n]}$.)

>From the general expression (\ref{7.6b}), it is clear that the pattern
(\ref{0.5}) emerges in a generic model (\ref{4.1}) as well. In particular,
owing to the pattern of eq. (\ref{7.6c}), the large $N$ asymptotics
$-ln[Q_{n}(\Gamma)]=n\tilde{\sigma}_{0}(\{\tilde{b}_{k}\})(1+O(1/N))$
is consistently provided  by the $\hat{T}_{\{p\}}=\hat{1}$ term
of (\ref{7.6b}). For $T_{\{p\}}\neq{\hat{1}}$, the leading $l=0$ term in
eq. (\ref{7.6c}) describes the branch-point canonically parametrized (see e.g.
\cite{Moore}) by $T_{\{p\}}$. The latter point decreases the associated Euler
character by $\sum_{k} (k-1)p_{k}$ which matches (akin to the pattern
(\ref{7.4}) of $\Omega_{n}$) with power of the {\it 1/N} factor assigned in
(\ref{7.6c}) to $\hat{T}_{\{p\}}$ . As it is discussed in Appendix E (where
the earlier results \cite{Ganor,Moore} are summarized and reformulated), the
$l\geq{1}$ terms can be reinterpreted as the movable subsurfaces (of various
topologies) collapsed to a point.

As for the factor $P_{n}^{(N)}(T_{\{p\}})$, inherited from the decomposition
(\ref{7.11xbx}) of the projector $P_{n}^{(N)}$, its dependence on {\it N}
is not particularly suitable for a manifest $1/N$ expansion like (\ref{0.5}).
It calls for a nontrivial resummation which would reproduce the
well-defined large $N$ {\it SC} series (\ref{0.5}) (obtained by Gross and
Taylor without resort to (\ref{7.8})). The latter pattern effectively
eliminates $P_{n}^{(N)}(T_{\{p\}})$ at the expense of working with
the $S(n^{+})\otimes S(n^{-})$ double-representation, $B(n)\rightarrow
{B(\{n^{\pm}\})}$, that refers to the two coupled sectors of the
opposite worldsheet orientation. The prescription to reconstruct
$B(\{n^{\pm}\})$ from the amplitude like (\ref{7.1}) is quite simple:
all the involved $S(n)$-structures are traded
for their $S(n^{+})\otimes S(n^{-})$ counterparts
\be
\delta_{n}(..)\rightarrow{\delta_{n^{+}\times n^{-}}(..)}~~,~~
N^{n}\Omega_{n}\rightarrow{N^{n^{+}+n^{-}}\Omega_{n^{+},n^{-}}}~~,~~
\sigma_{\rho}\rightarrow{\sigma^{+}_{\rho}\otimes \sigma^{-}_{\rho}},
\label{7.8b}
\ee
\be
Q_{n}\rightarrow{Q_{n^{+},n^{-}}=
e^{-\frac{\lambda}{2}(n^{+}+n^{-}-
((n^{+})^{2}+(n^{-})^{2}-2n^{+}n^{-})/N^2)}}
e^{-{\frac{\lambda}{N}(\hat{T}^{(n^{+})}_{2}+\hat{T}^{(n^{-})}_{2})}},
\label{7.8c}
\ee
where eq. (\ref{7.8c}) imples the Heat-Kernal case (\ref{4.1b}) and can be
generalized to a generic model (\ref{4.1}). (The definition and
interpretation of $\Omega_{n^{+},n^{-}}$ and other ingredients in eqs.
(\ref{7.8b}),(\ref{7.8c}) can be found in \cite{Gr&Tayl}.)

\subsection{Construction of the branched covering spaces.}

In the remaining subsections, we discuss the major issues
related to the effective enumeration of the mappings (\ref{7.11bb}) and
their automorphisms.
Let us proceed with an explicit algorithm which, given the symmetric group
data in the r.h.s. of (\ref{7.12}), reconstructs the topology of the
Riemann surfaces $\tilde{M}(\{p\},n,i)$ in the l.h.s. of (\ref{7.12}). As the
elements $\hat{T}_{\{p\}},~(\hat{T}_{2})^{i}$ are associated \cite{Ezell} to
the branch points (BPs), it is convenient to start with the simpler case of
the topological covering spaces (without BP's singularities) removing 
the latter elements. The full branched covering spaces (BCSs) can be
reproduced  reintroducing the BPs onto the corresponding covering spaces
(CSs).

As for a particular {\it n}-sheet CS $\tilde{M}$ of a given $2d$ surface $M$
(the 2-torus in what follows), it can be composed with the help of the
cutting-gluing rules borrowed from the constructive topology. To begin with,
cut a 2-torus {\it M} along the two uncontractible cycles $\alpha(\rho)$
trading the latter for the pairs of edges
${\alpha(\rho)\otimes{\beta(\rho)}},~
\rho=\mu,\nu$. It makes {\it M} into a rectangular $H_{\mu\nu}$ with the
boundary edge-path represented as
$\alpha(\mu)\alpha(\nu)\beta^{-1}(\mu)\beta^{-1}(\nu)$. Then, consider the
{\it trivial} covering $\tilde{H}_{\mu\nu}=H_{\mu\nu}\otimes \Upsilon_{n}$
(where $\Upsilon_{n}=\{1,...,n\}$) of $H_{\mu\nu}$ by
$n$ copies of this rectagular with the boundary edge-paths given by 
$\alpha(\mu_{k})\alpha(\nu_{k})\beta^{-1}(\mu_{k})\beta^{-1}(\nu_{k}),~
k=1,...,n$. Perform the set of the {\it pairwise} reidentifications of the
involved edges
\be
\alpha(\rho_{k})=\beta(\rho_{\sigma_{\rho}(k)})~~~~~;~~~~~
\sigma_{\rho}~:~~k\rightarrow{\sigma_{\rho}(k)}~~~,~~~
k\in{\Upsilon_{n}}~,
\label{7.16}
\ee
where $\sigma_{\mu}\equiv{\sigma_{1}},~\sigma_{\nu}
\equiv{\sigma_{2}}$ are supposed to satisfy the $\delta_{n}$-constraint
(\ref{7.12}) (with the excluded contribution of
$\hat{T}_{\{p\}}(\hat{T}_{2})^{i}$).

Evidently, the two sets
(\ref{7.16}) of the reidentificatrions can be concisely represented as the two
{\it closed} (i.e. without branch end-points) {\it branch cuts}
$\varpi_{\rho}$ of the Riemann surface $\tilde{M}(\{\sigma_{\rho}\})$ with
{\it n} sheets. According to the pattern (\ref{7.16}), each connected
component of $\tilde{M}(\{\sigma_{\rho}\})$ has the topology of 2-torus. In
compliance with (\ref{0.5}), it matches with the {\it N}-independence of the
argument of the $\delta_{n}$-function (\ref{7.8}) taking place after the
exclusion of $P_{n}^{(N)} exp[-\lambda\hat{T}^{(n)}_{2}/N]$.

To reintroduce the branch points (encoded in the
$\hat{T}_{\{p\}}(\hat{T}_{2})^{i}$ factor of (\ref{7.12})), recall that
each admissible BP is the end-point $q_{k}$ of the associated branch cut 
$\varpi^{(k)}$ which should be included additionally to
the closed cuts $\varpi_{\rho}$ of the CS $\tilde{M}(\{\sigma_{\rho}\})$.
To implement $\varpi^{(k)}$, we first cut $\tilde{M}(\{\sigma_{\rho}\})$
along the support of $\varpi^{(k)}$. (Both $\varpi^{(k)}$ and $\varpi_{\rho}$
are all supposed to terminate at a common base-point
$p=\alpha(\mu) \cap \alpha(\mu)$ of {\it M}.) Then, on the left and on the
right sides of each cut $\varpi^{(k)}$,  the resulting two copies of
the {\it n} new edges of the sheets are reidentified according to the
prescription (\ref{7.16}). The
only modification is that, instead of $\sigma_{\rho}$, one is to substitute
the appropriate permutations $\xi^{\{p\}}\in{T_{\{p\}}}$ and
$\tau^{(s)}\in{T_{2}}$ entering respectively
$\hat{T}_{\{p\}}$ and the $s$th $\hat{T}_{2}$-factor in the inner product
$(\hat{T}_{2})^{i}$. It completes the construction of the admissible
Riemann surfaces $\tilde{M}(\{p\},n,i)$ entering the l.h.s. of
(\ref{7.12}).

\subsection{The homomorphism of $\pi_{1}(M-\{q_{s}\}|p)$ into $S(n)$.}

To enumerate the equivalence classes of BCSs and justify
the asserted interpretation of $|C_{f}(\{p\})|$,
one is to employ the relation to the following 
{\it group homomorphism} \cite{Ezell}
where the first homotopy group $\pi_{1}(M-\{q_{s}\}\}|p)$
\be
{\bf \psi}~ :~~~
\pi_{1}(M-\{q_{s}\}|p)\rightarrow{S(n)}~,
\label{7.12bb}
\ee
is mapped into $S(n)$. (In eq. (\ref{7.12bb}), $M-\{q_{s}\}$ denotes
the base-surface {\it M} with the $i+1$ excluded points $q_{s}$, associated to
the branch points, and with the base-point {\it p}.)
Given an BCS encoded in the $\delta_{n}$-function (\ref{7.12}), choose a
set $\Upsilon_{n}=\{1,2,...,n\}$ to label the $n$ sheets (at the base-point
$p$).
Consider the lift \cite{Massey,Moore} of the closed paths in $M-\{q_{s}\}$
(defining $\pi_{1}(M-\{q_{s}\}|p)$) into the covering space
$\tilde{M}-\{f^{-1}(q_{s})\}$. Then, the (equivalence classes of the) paths
induce the permutations of the labels which determine the corresponding
$S(n)$ operators acting on $\Upsilon_{n}$.

To make (\ref{7.12bb}) explicit, let us first specify the pattern of the
first homotopy group. Consider a topological space $T-\{q_{1},...,q_{m}\}$
which, for our later
purposes, is allowed to be a (CW) $2d$ cell-complex $T$ ({\it not}
necessarily reduced to a $2d$ surface) with $m$ deleted points $q_{k}$.
Recall that in this case the group $\pi_{1}(T-\{q_{1},...,q_{m}\})$ (in what
follows we will everywhere omit the specification of the base-point {\it p})
can be represented as the following abstract group
\cite{Massey}. The generators of the latter group are associated to the
homotopy equivalence classes (HEC) of the {\it uncontractible} closed paths
based at a given point $p$ (supposed to be distinct from the set $\{q_{s}\}$).
In the case at hand, additionally to the generators $\alpha_{r},~r=1,...,P$
(corresponding to $P$ HECs of the uncontractible cycles of $T$), there are
extra $m$ generators $\gamma^{(s)},~s=1,...,m,$ which refer to the HECs of
closed paths encircling a single deleted point $q_{s}$. Finally, there exists
a constructive algorithm to find the complete set of $K\in{\bf Z_{\geq{1}}}$
{\it relations} $\{F_{l}(\{\alpha_{r},\gamma^{(s)}\})=1\}_{l=1,...,K}$
that completes the intermediate mapping of the
$\pi_{1}(T-\{q_{1},...,q_{m}\})$ generators into the abstract group.

Returning to the case of a genus $g$ $2d$ surface $T=M_{g}$, there is a
single relation (with $P=2g,~[\alpha_{i},\alpha_{k}]
\equiv{\alpha_{i}\alpha_{k}\alpha^{-1}_{i}\alpha^{-1}_{k}}$) defining
$\pi_{1}(M_{g}-\{q_{s}\})$
\be
F(\{\alpha_{r},\gamma^{(s)}\})=\left(\prod_{j=1}^{g}~
[\alpha_{j},\alpha_{g+j}]~
\prod_{s=1}^{m}\gamma^{(s)}\right)=1~.
\label{7.12c}
\ee
Comparing (\ref{7.12c}) with the pattern (\ref{7.8})/(\ref{7.12}),
one deduces the explicit form of the homomorphism (\ref{7.12bb})
(with the identification $g=1,~m=i+1$)
\be
{\bf \psi }~:~~\psi(\alpha_{\rho})={\sigma_{\rho}}~;~~
\psi(\gamma^{(1)})={\xi^{\{p\}}}~;~~\psi(\gamma^{(s)})={\tau^{(s-1)}}~,~
{s}\geq{2},
\label{7.12cc}
\ee
where $\xi^{\{p\}}\in{T_{\{p\}}}$ and $\tau^{(s-1)}\in{T_{2}}$ which enter
respectively $\hat{T}_{\{p\}}$ and the $(s-1)$th $\hat{T}_{2}$-factor
(in the inner product $(\hat{T}_{2})^{i}$).

\subsection{The symmetry factor.}

Given the homomorphisms (\ref{7.12cc}), one can readily
enumerate the equivalence classes $\tilde{M}(\{p\},n,i)$ of the BCSs
employing the following notion of
equivalence \cite{Gr&Tayl} of two homomorphisms ${\bf \psi}_{1}$ and
${\bf \psi}_{2}$. The latter are postulated to
belong to the same equivalence class if there exists some
$\eta\in{S(n)}$ so that
\be
\psi_{1}(\zeta)=\eta ~\psi_{2}(\zeta)~ \eta^{-1}~~,~~
\forall{\zeta}\in{\pi_{1}(M-\{q_{1},...,q_{m}\})}~~;~~\eta\in{S(n)}.
\label{7.12e}
\ee
Then, the basic theorem \cite{Ezell,Massey} of the topological coverings
ensures that the inequivalent homomorphisms (\ref{7.12cc}) are in one-to-one
correspondence with the associated homeomorphically distinct branched
covering spaces.

Finally, let $\kappa :~f\circ \kappa=f,$
denotes a particular automorphism of
the branched covering space $\tilde{M}$. Being restricted to the {\it n}-set
$\Upsilon_{n}=f^{-1}(\tilde{p})$ ($p\neq{q_{s}},~\forall{s}$), the group
of the automorphisms is isomorphic \cite{Massey} to 
(the conjugacy class of) the $S(n)$-subgroup $C_{f}(\{p\})$ that
induces conjugations (\ref{7.12e}) leaving {\it all} the images
$\psi_{1}(\zeta)$ {\it invariant}:
${\bf \psi}_{1}(\zeta)={\bf \psi}_{2}(\zeta),~\forall{\zeta},~
\forall{\eta(\kappa)}\in{C_{f}(\{p\})}$. In turn, it
justifies \cite{Gr&Tayl} the required interpretation of $|C_{f}(\{p\})|$.

\section{The stringy form of $B(\{n_{\mu\nu}\})$ in $D\geq{3}$.}

To rewrite the $D\geq{3}$ amplitude (\ref{7.7}) in the form generalizing the
$D=2$ stringy pattern (\ref{7.11})-(\ref{7.12}),
we first expand each factor
$Q_{n_{\mu\nu}}$ and select the $i_{\mu\nu}$th power
$(\hat{T}^{(n_{\mu\nu})}_{2})^{i_{\mu\nu}}$ (where the $k_{\mu\nu}$th
$\hat{T}^{(n_{\mu\nu})}_{2}$-factor in the latter product is supposed to be
defined via eq. (\ref{7.6}) in terms of
$\tau^{(k_{\mu\nu})}_{\mu\nu}\in{{T}^{(n_{\mu\nu})}_{2}},~
k_{\mu\nu}=1,...,i_{\mu\nu}$). Complementary, akin to eq. (\ref{7.6b}) one is
to decompose
\be
\Lambda_{n_{\phi}}^{(m_{\phi})}=
\sum_{T^{(n_{\phi})}_{\{p_{\phi}\}}\in{S(n_{\phi})}}
\Lambda^{(m_{\phi})}_{n_{\phi}}(T^{(n_{\phi})}_{\{p_{\phi}\}})~
\sum_{\xi^{\{p_{\phi}\}}\in{T^{(n_{\phi})}_{\{p_{\phi}\}}}}
\xi^{\{p_{\phi}\}}~,
\label{7.15}
\ee
and separate the contribution of a given $\hat{T}^{(n_{\phi})}_{\{p_{\phi}\}}
\equiv{\hat{T}_{\{p_{\phi}\}}}$
(defined by eq. (\ref{7.6b})).

Given the above expansions, one can prove  (see below) that the associated
building block of (\ref{7.7})
$$
\sum_{\{\sigma_{\rho}\}}
\delta_{n_{+}}(\hat{T}_{\{p_{+}\}}
\bigotimes_{\{\mu\nu\}}
\hat{T}_{\{p_{\mu\nu}\}}
\frac{(\hat{T}_{2}^{(n_{\mu\nu})})^{i_{\mu\nu}}}
{n_{\mu\nu}!}
\prod_{\{\rho\}}
(\sigma_{\rho} \otimes \hat{1}_{[\frac{n_{+}}{n_{\rho}}]})
\prod_{\{\mu\}}
(\sigma^{-1}_{\mu}\hat{T}_{\{p_{\mu}\}}\otimes
\hat{1}_{[\frac{n_{+}}{n_{\mu}}]}))=
$$
\be
=\sum_{\varphi\in{\tilde{M}(\{p_{\phi}\},\{n_{\phi}\},\{i_{\mu\nu}\})}}
\frac{1}{|C_{\varphi}(\{p_{\phi}\})|}
\label{7.17}
\ee
can be rewritten as the sum over the relevant $D\geq{2}$ mappings
(\ref{0.1}) to be reconstructed in the next subsection. Extending the $D=2$
theorem due to Gross and Taylor, each term of the latter sum is weighted by
the inverse number $|C_{\varphi}(\{p_{\phi}\})|$ of distinct automorphisms
associated to a given $\tilde{M}_{\varphi}
\equiv{\tilde{M}(\{p_{\phi}\},\{n_{\phi}\},\{i_{\mu\nu}\})}$.

As we will demonstrate, the involved into (\ref{7.17}) spaces
$\tilde{M}_{\varphi}$ can be viewed as the generalized Riemann
surfaces to be identified with the worldsheets of the {\it YM}-flux 'wrapped
around' the EK $2d$ cell-complex (\ref{0.3}).
Combining all the pieces together, the stringy reinterpretation of
(\ref{7.7}) essentially follows the steps discussed in the $D=2$ case. Namely,
leaving aside the
$P_{n_{\phi}}^{(N)}(T_{\{p_{\phi}\}})$ factors (\ref{7.1}), the
rest of the ingredients of (\ref{7.7}) readily fit in the consistent
$D\geq{3}$ extension (\ref{0.1}) of the Gross-Taylor stringy pattern
(\ref{0.5}). Indeed, by the same token as in the $D=2$ analysis,
the involved (movable or nonmovable) branch points, microscopic tubes, and
handles are weighted by the $1/N$- and $n_{\phi}$-dependent factors according
to their contribution to the Euler character of $\tilde{M}_{\varphi}$. The
latter 'local' matching is completed by eq. (\ref{7.17}) together with
the following 'global' matching. To see it,
let us remove temporarily the branch points and the collapsed subsurfaces via
the substitution 
\be
\Lambda_{n_{\phi}}^{(m_{\phi})}~
\rightarrow{~N^{n_{\phi}m_{\phi}}}~~~~~;~~~~~
Q_{n_{\mu\nu}}\rightarrow{exp[-\lambda n_{\mu\nu}/2]}~,
\label{7.17cb}
\ee
into eq. (\ref{7.7}). Then, the remaining $\lambda n_{\mu\nu}$-dependent
factor and the overall power $(1/N)^{\varepsilon}$ (inherited from eq.
(\ref{7.7}))
\be
\varepsilon={\sum\limits_{\mu\nu=1}^{D(D-1)/2} n_{\mu\nu}-
\sum\limits_{\rho=1}^{D} n_{\rho}+n_{+}}=0
~~~;~~~
2n_{+}=\sum_{\{\rho\}} n_{\rho}=2\sum_{\{\mu\nu\}} n_{\mu\nu},
\label{7.17b}
\ee
match with the area and the 2-tora topology (revealed below) of each
connected component of $\tilde{M}_{\varphi}$ corresponding to
the deformation (\ref{7.17cb}).

By the same token as in the $D=2$ case, the manifest stringy representation of
the full amplitude (\ref{7.7}) (including the
$P_{n_{\phi}}^{(N)}(T_{\{p_{\phi}\}})$-factors) calls for
a resummation into the appropriate large $N$ {\it SC} series. Presumably, it
eliminates these factors trading each remaining in (\ref{7.7})
$S(n_{\phi})$-operator for its $S(n^{+}_{\phi})\otimes S(n^{-}_{\phi})$
descedant. It is suggestive that the short cut way for
the latter reformulation is provided by the direct extention of the $D=2$
prescription (\ref{7.8b}),(\ref{7.8c}): one is to 
substitute the labels $n,n^{+},n^{-}$ 
by their properly associated $\phi$-dependent counterparts
$n_{\phi},n^{+}_{\phi},n^{-}_{\phi}$ (where $n_{\phi},~\phi\in{\{\mu\nu\},
\{\rho\},+},$ labels the relevant $S(n_{\phi})$-structures). Note also that, 
similarly to
the $D=2$ case, a generic $D\geq{3}$ model (\ref{4.1}) complies with the
pattern (\ref{0.5}) as well.

In conclusion, we remark that the amplitude (\ref{7.7})
exhibits purely topological assignement (see Section 6.2 for more details) of
the $S(n_{\phi})$-structures after the formal deformation
$\{\Lambda_{n_{\phi}}^{(m_{\phi})}\rightarrow
{(N^{n_{\phi}}\Omega_{n_{\phi}})^{m_{\phi}}}\},~
\{Q_{n_{\mu\nu}}\rightarrow{1}\}$. In particular, consider
the powers $m_{\phi}=\pm 1$ in which the relevant
$S(n_{\phi})$-twists $\Omega_{n_{\phi}}$ enter the
$\Lambda^{(m_{\phi})}_{n_{\phi}}$ factors (\ref{7.1}): $\{m_{\mu\nu}=1\},~
\{~m_{\rho}=-1\},~~m_{+}=1$.
The latter are equal to the weights in the famous
formula computing the Euler characteristic of a given cell-complex $T$: 
$2-2G_{T}=n_{p}-n_{l}+n_{s}=n_{p}m_{p}+n_{l}m_{l}+n_{s}m_{s}$,
where $n_{p},~n_{l},~n_{s}$ are the total numbers of respectively plaquettes,
links, and sites of $T$. (The earlier heuristic arguments (in the case of the
single chiral sector), consistent with the above pattern 
of the $\Omega_{n_{\phi}}$ assignement, can be found in \cite{Ramg}).

The topological nature of the considered defomation of the amplitudes like
(\ref{7.17}) can be made transparent by the conjecture
building on the $D=2$ observation \cite{Moore}.
One may expect that, after this deformation, the TEK partition
function
$\tilde{X}_{D}$ yields the generating functional for the orbifold Euler
characters of the Hurwitz-like spaces associated to the following set. The
latter includes all the generalized branched covering spaces
$\tilde{M}_{\varphi}$ (of the EK $2d$ cell-complex (\ref{0.3})) corresponding
in eq. (\ref{7.17}) to the considered
deformation $\{P_{n_{\phi}}^{(N)}\rightarrow{1}\}$. (For a close but somewhat
distinct earlier conjecture, see \cite{Ramg}.) Similar proposal can be made
for the (weaker) 'topological' deformation $\{Q_{n_{\mu\nu}}\rightarrow{1}\}$ 
of the PF $\tilde{X}_{D}$ reintroducing the $P_{n_{\phi}}^{(N)}$ factors.

\subsection{The generalized covering spaces of $T_{EK}-\{q_{s}\}$.}

Given the data in the argument of the $\delta_{n_{+}}$-function in the l.h.s
of eq. (\ref{7.17}), our aim is to reconstruct the
associated topological spaces $\tilde{M}_{EK}\equiv{\tilde{M}_{\varphi}}$ (i.e.
the mappings (\ref{0.1})) which are summed over in the r.h.s. of (\ref{7.17}).
This problem, being {\it inverse} to what is usually considered in the
framework of algebraic topology \cite{Massey}, will be resolved through the
sequence of steps which appropriately generalizes the $D=2$ analysis
of the previous section. It is noteworthy that the spaces $\tilde{M}_{EK}$ do
{\it not} coincide with the canonical branched covering spaces (BCSs) of the
TEK $2d$ cell-complex $T_{EK}$. To say the least, a generic canonical BCS of
$T_{EK}$ is again a $2d$ cell-complex (with a number of branch points) but not
a $2d$ surface like in the asserted mappings (\ref{0.1}). As we will see, the
basic amplitude (\ref{7.17}) indeed complies with (\ref{0.1}) and refers to
the {\it novel} class of the associated
to $T_{EK}$ spaces $\tilde{M}_{EK}$ to be called the generalized branched
covering spaces (GBCSs) of $T_{EK}$.

Similarly to the $D=2$ analysis, we start with the simpler case of the
generalized covering spaces (GCSs) of $T_{EK}$ corresponding to the
deformation (\ref{7.17cb}). Being specified by the immersions (\ref{0.1})
without the (branch points' and collapsed subsurfaces') singularities,
the GCSs can be reconstructed generalizing the $D=2$ surgery-construction of
the covering spaces. Take $D(D-1)/2$
$\mu\nu$-rectangulars $H_{\mu\nu}$ representing the 'decompactified'
plaquettes of the EK base-lattice (\ref{0.3}). Then, one is to
begin with trivial ($n_{\mu\nu}$-sheet) coverings
$\tilde{H}_{\mu\nu}=H_{\mu\nu}\otimes \Upsilon_{n_{\mu\nu}}$ of
$H_{\mu\nu}$ with the
edge paths $\otimes_{q} \alpha(\mu_{q})
\alpha(\nu_{q})\beta^{-1}(\mu_{q})
\beta^{-1}(\nu_{q}),~q=1,...,n_{\mu\nu}$. To reproduce the effect of
the $\sigma_{\rho}$-permutations in (\ref{7.17}), first let us denote
by $\{\alpha(\rho_{k}),~k=1,...,n_{\rho}=
\sum_{\nu\neq{\rho}}^{D-1} n_{\rho\nu}\}$ the $\rho$-set of the 
$\alpha(\rho_{q})$ edges (collected from the $(D-1)$ different
$\rho\nu$-plaquettes) ordered in accordance with the pattern
(\ref{5.13b}) of the $S(n_{\rho})$ basis $|I^{(\pm)}_{n(\rho)}>$. Similarly,
we introduce the sets $\{\beta(\rho_{k}),~k=1,...,n_{\rho}\}$.

To reconstruct the associated to (\ref{7.7}) GCS of $T_{EK}$, at each
particular $\rho$-link one is to perform the pairwise
$\sigma_{\rho}$-identifications of $\alpha(\rho_{k}),\beta(\rho_{k})$
according to the prescription (\ref{7.16}). In this way, we have constructed
appropriate conglomerate of the generalized Riemann surfaces
$\tilde{M}(\{\sigma_{\rho}\})$
(with the total number $D$ of the closed branch cuts) which are wrapped around
$T_{EK}$ in compliance with the mapping (\ref{0.1}). Evaluating the
Euler characteristic $\varepsilon$ of $\tilde{M}(\{\sigma_{\rho}\})$ in
accordance with (\ref{7.16}), each connected component of the latter surface
reveals the topology of a 2-tora. In turn, it matches with the overall power
(\ref{7.17b}) of the $1/N$ factor 
which is inherited via (\ref{7.17cb}) from the product (\ref{7.7}) of
the $\Lambda_{n_{\phi}}^{(m_{\phi})}$ factors.

Given the GCSs $\tilde{M}(\{\sigma_{\rho}\})$ of $T_{EK}$, the branch points
(encoded in (\ref{7.17}) through
the operators $\hat{T}_{\{p_{\mu\nu}\}}
(\hat{T}_{2}^{(n_{\mu\nu})})^{i_{\mu\nu}}$) are
reintroduced in essentially the same way as we did for the $D=2$ case.
The only subtlety is that, to
make the employed cutting-gluing rules well-defined, it is convenient to view
the EK cell-complex $T_{EK}$ as the homotopy {\it retract} of a 'less
singular' $2d$ complex $T'_{EK}$ (possessing by construction the {\it same}
fundamental group $\pi_{1}(T_{EK})=\pi_{1}(T'_{EK})$). We refer the reader
to Appendix C for the details, and now proceed with the $D\geq{3}$
generalization of the $D=2$ Gross-Taylor theorem concerning the
interpretation of the symmetry factor entering the $D\geq{2}$ stringy
amplitudes like (\ref{0.5}),(\ref{7.17}).

\subsection{The homomorphism of $\pi_{1}(T_{EK}-\{q_{s}\})$ into $S(n_{+})$.}

To effectively enumerate the constructed GBCSs of $T_{EK}$ and their
automorphisms, we first reconstruct what kind of homomorphism like
(\ref{7.12bb}) is encoded in the $D\geq{3}$ pattern (\ref{7.17}). It will
provide with the precise mapping of the fundamental group
$\pi_{1}(T_{EK}-\{q_{s}\})$ (of $T_{EK}$
with a number of deleted points $\{q_{s}\}$ associated to the branch points)
into the enveloping $S(n_{+})$ group.

Observe first that the abstract group representation of $\pi_{1}(T_{EK})$ is
defined, according to (\ref{0.3}), in terms of the $D$ generators
$\alpha_{\rho}$ corresponding to the uncontractible cycles (i.e. the
compactified $\rho$-links) of $T_{EK}$. Having excluded from $T_{EK}$ a set
of points $\{q_{s}\}$, it is
convenient to recollect them into the three varieties of the $\phi$-subsets
$\{q_{s}\}=\cup_{\phi}\{q^{(\phi)}_{k_{\phi}}\},~
\phi\in{\{\mu\nu\},\{\rho\},+},~k_{\phi}=1,...,b_{\phi},$ belonging
respectively to the {\it interior} of the $\mu\nu$-plaquette, to the interior
of the $\rho$-link, and to the single site of $T_{EK}$.
The set of the $\pi_{1}(T_{EK}-\{q_{s}\})$ generators includes
(additionally to the subset $\{\alpha_{\rho}\}$ inherited from
$\pi_{1}(T_{EK})$) $\gamma^{(\phi)}_{k_{\phi}}$ associated
to (the equivalence classes of) the closed paths encircling a single
deleted point $q^{(\phi)}_{k_{\phi}}$. The representation of
$\pi_{1}(T_{EK}-\{q_{s}\})$ is then completed by 
the $D(D-1)/2$ relations 
\be
\left( [\alpha_{\mu},\alpha_{\nu}]~~\cdot
\prod_{\phi=\mu\nu,\mu,\nu,+}~~\prod_{k_{\phi}=1}^{b_{\phi}}
\gamma^{(\phi)}_{k_{\phi}} \right)=1~~~;~~~~\mu\nu=1,...,{D(D-1)}/{2},
\label{7.12ccc}
\ee
each of which can be viewed as the $\mu\nu$-'copy' of (\ref{7.12c}).

Comparing the argument of the $\delta_{n_{+}}$-function in the l.h.s.
of eq. (\ref{7.17}) with the pattern of (\ref{7.12ccc}), it is straightforward
to write down the precise form of the relevant $D\geq{2}$ homomorphism
generalizing (\ref{7.12bb}):
\be
{\bf \psi } :~\psi(\alpha_{\rho})=\sigma_{\rho}~;~~
\psi(\gamma^{(\phi)}_{1})=\xi^{\{p_{\phi}\}}~;~~
\psi(\gamma^{(\mu\nu)}_{k_{\mu\nu}})=\tau^{(k_{\mu\nu}-1)}_{\mu\nu}~,~
{k_{\mu\nu}}\geq{2},
\label{7.12cb}
\ee
where $\xi^{\{p_{\phi}\}}\in{{T}_{n_{\phi}}},~
\tau^{(k_{\mu\nu}-1)}_{\mu\nu}\in{{T}_{2}^{(n_{\mu\nu})}}$. Note that the
three $\phi$-varieties of the generators $\gamma^{(\phi)}_{1}$
match with the three species of $\hat{T}_{n_{\phi}}$ in eq.
(\ref{7.17}), while $\gamma^{(\mu\nu)}_{k_{\mu\nu}}$ represents the
$k_{\mu\nu}$th $\hat{T}_{2}^{(n_{\mu\nu})}$-factor in the product
$(\hat{T}_{2}^{(n_{\mu\nu})})^{i_{\mu\nu}}$. In particular, eq. (\ref{7.12cb})
implies that in eq. (\ref{7.12ccc}) $b_{\mu\nu}=i_{\mu\nu}+1$ while
$b_{+}=b_{\rho}=1,~\forall{\rho}$.

\subsubsection{The symmetry factor.}

To deduce the announced interpretation of $|C_{\varphi}(\{p_{\phi}\})|$
(entering the r.h.s. of eq. (\ref{7.17})),
let us start with the following observation. The summation in the l.h.s of
(\ref{7.17}) can be viewed as the sum over the associated
homomorphisms (\ref{7.12cb}) constrained by the
condition: $\xi^{\{p_{\phi}\}}\in{T_{\{p_{\phi}\}}}$ and
$\tau^{(k_{\mu\nu}-1)}_{\mu\nu}\in{{T}^{(n_{\mu\nu})}_{2}}$.
Therefore, one is to identify the proper equivalence classes of the homomorphisms
(\ref{7.12cb}) to parametrize uniquely the topologically inequivalent
spaces $\tilde{M}_{\varphi}$ constructed in Section 6.1.

We assert that the two homomorphisms (\ref{7.12cb}), $\psi_{1}$ and
$\psi_{2}$, are {\it equivalent} (i.e. the corresponding generalized branched
coverings $\tilde{M}_{\varphi}$ are homeomorphic) if and only if there exists
some $\eta\in{\otimes_{\mu\nu}S(n_{\mu\nu})}$ so that
\be
\psi_{1}(\zeta)=\eta \psi_{2}(\zeta) \eta^{-1}~~,~~
\forall{\zeta}\in{\pi_{1}(T'_{EK}-\{q_{s}\})}~~;~~
\eta\in{\otimes_{\mu\nu} S(n_{\mu\nu})},
\label{7.12ex}
\ee
where the basis (\ref{5.6c}) for $\eta\in{S(n_{+})}$ is implied. For a
preliminary orientation, one observes that the conjugations (\ref{7.12ex})
are induced by the $\otimes_{\mu\nu} S(n_{\mu\nu})$ permutations of
the sheets (of $\tilde{M}_{\varphi}$) {\it separately} within each of the
$D(D-1)/2$ distinct $n_{\mu\nu}$-subsets which covers (see Section 6.1) a
given $\mu\nu$-tora $E_{\mu\nu}$ combined into $T'_{EK}$. (Complementary, 
$\otimes_{\mu\nu} S(n_{\mu\nu})$ is the {\it largest} subgroup of $S(n_{+})$
providing with the conjugations leaving the argument of the
$\delta_{n_{+}}$-function
in eq. (\ref{7.17}) invariant.) We refer to Appendix D for the justification
of the notion (\ref{7.12ex}) of the equivalence and now simply deduce its
consequences. Consider a particular branched covering $\tilde{M}_{\varphi}$,
and let $C_{\varphi}(\{p_{\phi}\})$ is the group of the inequivalent
automorphisms $\kappa$ of $\tilde{M}_{\varphi}$: $\varphi\circ \kappa=
\varphi$. Take the restriction of a given $\kappa$ to the $\Upsilon_{n_{+}}$
space of $\{\varphi^{-1}(p)\})$. Akin to the $D=2$ case, one shows that
$C_{\varphi}(\{p_{\phi}\})$ is isomorphic to the conjugacy class
(with respect to (\ref{7.12ex})) of the following subgroup of
$\otimes_{\mu\nu} S(n_{\mu\nu})$. The latter is associated to such subset
of conjugations (\ref{7.12ex}) that leave all $\psi_{1}(\zeta)$ invariant:
$\psi_{1}(\zeta)=\psi_{2}(\zeta),~\forall{\zeta},~
\forall{\eta(\kappa)\in C_{\varphi}(\{p_{\phi}\})}$.
In sum, there are exactly
$(\otimes_{\mu\nu}n_{\mu\nu}!)/|C_{\varphi}(\{p_{\phi}\})|$ distinct 
homomorphisms $\{\psi_{k}\}$ associated to the same topology-type of 
$\tilde{M}_{\varphi}$ that justifies the basic identity (\ref{7.17}).

\section{The area-preserving homeomorphisms.}

On a lattice, our $D\geq{3}$ stringy proposal can be viewed as the
confluence of the Wilson's string-like reformulation \cite{Wils} of
the ({\it finite} {\it N}) {\it SC} series with the power of the large {\it N}
expansion. The feature, which sharply
distinguishes the $D\geq{3}$ Gauge String from the earlier $D\geq{3}$
proposals \cite{Kaz&Zub,Kost1} in this direction, is the asserted invariance
of the weights $w[\tilde{M}]$ under certain {\it continuous} group of the
area-preserving worldsheet homeomorphisms. In $D\geq{3}$, the latter symmetry
is encoded in the following $D\geq{3}$ 'descedant' inherited from the
$D=2$ renormgroup ({\it RG}) invariance of the $YM_{2}$ systems (\ref{4.1}).

Recall first that a $D\geq{3}$ $YM_{D}$ theory (\ref{4.1})
(having some lattice ${\bf L^{D}}$ as the base-space) can be equally viewed
as the $YM$ theory being defined on the $2d$
skeleton ${\bf T^{D}}$ of ${\bf L^{D}}$ represented by the associated
$2d$ cell-complex. Consider the partition function (PF) of (\ref{4.1})
on ${\bf T^{D}}=\cup_{k} E_{k}$ composed from the associated $2d$ surfaces
$E_{k}$ (of the areas $A_{k}$ and with certain boundaries) according to the
corresponding incidence-matrix \cite{Dr&Zub}.
Akin to the $D=2$ case, this PF is
invariant under subdivisions of ${\bf T^{D}}$ (preserving the total areas
$A_{k}$) so that $E_{k}$ can be made into the associated smooth $2d$
manifolds $M_{k}$ (with boundaries) combined into a $2d$ cell-complex
${\bf \tilde{T}^{D}}=\cup_{k} M_{k}$ (homeomorphic to ${\bf T^{D}}$).
Therefore, refining the discretization, the $YM_{D}$ lattice theory
(\ref{4.1}) can be adjusted to merge with the following
continuous system. To implement the latter, we first put
the associated to (\ref{4.1}) continuous $YM_{2}$ theory (\ref{0.1ee})
(keeping free boundary conditions) on each $M_{k}$ . Then, one is to
average over the gauge fields, 'living' on the boundaries of $\{M_{k}\}$,
in compliance with the associated to ${\bf \tilde{T}^{D}}$
{\it incidence-matrix}.

Similarly to the $D=2$ case (see Section 1.1A), on the side of the proposed
$D\geq{3}$ Gauge String, the above relation to the continuous $YM_{2}$ system
ensures
the required invariance of the lattice string weights (entering the amplitudes
defined by the $D\geq{3}$ extension (\ref{0.1}) of (\ref{0.5})). Indeed,
on the one hand, both the parity $P_{\varphi}$ and the symmetry factor
$|C_{\varphi}|$ depend only on the {\it topology} of the worldsheet
$\tilde{M}(T)$ and on the one of the corresponding taget-space
$T=\cup_{k} E_{k}$.
(Compare it with \cite{Kaz&Zub,Kost1}
where the singular lattice {\it geometry} of {\it T} is 'built into' the
curvature-defects.) On the other hand, the sum over the mappings $\varphi$
also supports the required invariance of the set of relevant $\tilde{M}(T)$.
Indeed, the multiple
{\it integrals} (rather than discrete sums as in \cite{Kaz&Zub,Kost1}) over
all admissible positions of the movable branch points (and certain collapsed
subsurfaces) span the {\it entire} interior of the discretized
$2d$ surfaces $E_{k}$ combined into {\it T}. The remaining 'discrete'
contributions into $\int d\varphi$ refer to the purely topological
assignement of the nonmovable singularities {\it anywhere} in the interiors
of the appropriate subspaces of {\it T}.
Altogether, the relevant group of the worldsheet homeomorphisms
continuously translates the positions of the admissible singularities
of the map (\ref{0.1}) within the interiors of the associated subspaces of
{\it T}.

To be more specific, we compare the weights $w[\tilde{M}]$ associated to
the $YM$ system (\ref{4.1}) defined on a generic base-lattice $F_{EK}$
homeomorphic to the 'elementary' {\it EK} $2d$ cell-complex (\ref{0.3}). Let
the total area (measured in the dimensionless units) of a given torus
$E_{\mu\nu}$ is equal to $A_{\mu\nu}$. Combining the above general arguments
with the analysis of Sections 3 and 4, the general weights on
$F_{EK}$ can be deduced from the basic $T_{EK}$ amplitude (\ref{7.7})
trading in each $Q_{n_{\mu\nu}}$-factor (\ref{7.6}) the coupling constant
$\lambda$ for $\lambda A_{\mu\nu}$. In particular,  
the $(\Omega_{n_{\phi}})^{m_{\phi}}$-, $\sigma_{\rho}$-twist assignement is
indeed purely topological: the corresponding conglomerates of the
nonmovable branch points (and collapsed subsurfaces) can be placed anywhere
(but without summation over positions) in the interior of the associated
macroscopic $\phi$-,$\rho$-cell of $F_{EK}$.

\section{Smooth Gauge Strings vs. lattice ones.}

It is clear that the method, we have developed in Sections 3-6
on the example of the TEK model (\ref{2.1}), can be extended for the $YM_{D}$
theories (\ref{4.1}) defined on a generic regular subspace {\it T} of the
$2d$ skeleton of the {\it D}-dimensional base-lattice ${\bf L^{D}}$. We
defer the analysis of the generic weights (\ref{0.5c}) for a separate paper
and now simply stress those implications of our present results which are
{\it T}-independent and common for both the lattice and smooth realizations of
the Gauge String. In this way, one actually decodes all the major peculiar
features of the continuous theory of the smooth {\it YM}-fluxes which are
novel compared to the conventional paradigm of the $D\geq{3}$ 'fundamental'
strings.

To begin with, one observes that the relevant smooth mapping (see e.g. eq.
(\ref{0.1}) for the worldsheets $\tilde{M}$ without boundaries) are allowed
to have singularities which usually are not included into the $D\geq{3}$
string-pattern (where the maps (\ref{0.1}) are restricted to be sheer
immersions). Complementary, certain class of the selfintersecting
worldsheets $\tilde{M}$ is endowed in eq. (\ref{0.5c}) with the extra factors
$J[\tilde{M}(T)|\{\tilde{b}_{k}\}]\neq{1}$ that does not have a
direct counterpart in the known $D\geq{3}$ string theories. In turn, it is
the extra $J[..]$-weights which encode the data sufficient to reconstruct
the associated continuous $YM_{D}$ model (\ref{0.1ev})/(\ref{0.1ee}).
As for a single $D\geq{3}$ Wilson loop average,
the major novel ingredient is due to the various movable branch points 
and the collapsed $2d$ subsurfaces
(assigned with the $\{\tilde{b}_{k}\}$-dependent weights 
(\ref{0.5cx})). Their positions can be viewed as the zero
modes associated to the minimal surface contribution 
(provided the latter properly selfintersects) resulting in the extra
area-dependence in the preexponent of $<W_{C}>$. This pattern generalizes 
the one of the $D=2$
loop-averages \cite{Gr&Tayl,Kost1}.

Next, let us demonstrate that the total contribution of the worldsheets
$\tilde{M}$ (and, consequently, of the taget-spaces {\it T}) with
the backtrackings vanishes which matches with the absence of foldings in the
$D=2$ Gross-Taylor representation (\ref{0.5}). To make this feature manifest, we
first recall that the relevant for (\ref{7.17}) maps (\ref{0.1}) (associated
to the {\it SU(N)} TEK model) do not include any foldings of the worldsheets
$\tilde{M}$ wrapped around $T_{EK}$. On
the other hand, one could equally start with the large {\it N} {\it U(N)} TEK
model where the proposed technique would result in all kinds of foldings
'covering' various conglomerates of the plaquettes (with possibly different
$\mu\nu$-orientation). The comparison of these two complementary large $N$
patterns justifies the above assertion.

On the side of the gauge theory defined on the standard cubic lattice
${\bf L^{D}}$, the TEK foldings are associated to the more general
backtrackings of $\tilde{M}$. To see it, let us call
worldsheets $\tilde{M}(T)$ (or taget-spaces {\it T}) without any
backtrackings regular. Then, to any regular taget $T^{(r)}$ one can associate
the variety of taget-spaces creating all kinds of the backtracking (bounding
a zero 3-volume) with the support {\it not} necessarily belonging the
original regular taget $T^{(r)}$ (in contradistinction to the foldings
discussed in \cite{Kost1}).

The irrelevance of the backtrackings is intimately related to the invariance
(in gauge theories) of the Wilson loop averages $<W_{C}>$ under the zig-zag
backtrackings of the boundary contour {\it C} that is in sharp contrast with
the situation in the conventional Nambu-Goto string and most of its existing
generalizations. (Recently Polyakov \cite{Polyak2} advocated the latter
invariance as the crucial feature of the strings dual to gauge theories.)
In the limit $N\rightarrow{\infty}$, the zig-zag symmetry of $<W_{C}>$ can be
made manifest confronting
as previously the two large {\it N} formulations of the open Gauge String:
the $U(N)$ one with a given backtracking contour {\it C} should be compared
with the $SU(N)$ one with $\tilde{C}$ obtained contracting zig-zags of $C$.

\subsection{Correspondence with the {\it WC} Feynman diagrams.}

Let us reveal the {\it WC}/{\it SC}
correspondence between the continuous $YM_{D}$ models (\ref{0.1ee}) in the
{\it WC} phase and the associated smooth Gauge String in the
{\it SC} regime. According to Section 7, once the
backtrackings are absent, the relevant weights $w[\tilde{M}(T)]$ can be
derived by putting the $YM_{2}$ system (\ref{0.1ee}) onto a given
$2d$ cell-complex (e.g. $2d$ surface) $T^{(r)}=\cup_{k} M_{k}$ where the
$2d$ surfaces $M_{k}$
are piecewise smooth. In the large {\it N} {\it SC} regime,
the latter $YM$ system is represented by the conglomerates of the
worldsheets properly wrapped around ${\it T^{(r)}}$. On the other hand, in the
large {\it N} {\it WC} regime the {\it WC} perturbation theory represents 
our system through the set the Feynman diagrams. Employing the
standard path integral representation of the propagator of the free particle 
(in a curved
space), the diagrams are visualized as the fishnet (of the gluonic
trajectories) appropriately 'wrapped' around the
same $2d$ complex ${\it T^{(r)}}$. The crucial observation is
that, when ${\it T^{(r)}}$ is viewed as being embedded into ${\bf R^{D}}$, the
latter fishnet can be reinterpreted as the specific contribution
of the {\it WC} perturbation theory in the
{\it D}-dimensional continuous $YM_{D}$ model (\ref{0.1ee}) (uniquely
associated to the chosen $YM_{2}$ via eq. (\ref{0.1ev})). To select this
contribution on the $YM_{D}$ side, not
only the gluonic trajectories in ${\bf R^{D}}$ should be constrained to have 
the space-time support on ${\it T^{(r)}}$ but also {\it each} colour
$a$-component of the associated gluonic
strength-tensor  $F^{a}_{\mu\nu}({\bf z}),~
{\bf z}\in {\it T^{(r)}},$ as the Lorentz (or $O(D)$) tensor should belong to the 
tangent space of ${\it T^{(r)}}$ at ${\bf z}$. (To circumvent
gauge-fixing, tricky to explicitly match between the $T^{(r)}$- and standard
formulations, one is to introduce an infinitesimally small mass term for the
gauge field and then perform the above comparison.)

\subsection{Suppression of the selfintersections.}

At this step, it is appropriate to discuss a number of crucial
simplifications inherent in the dynamics of the considered $D\geq{3}$
continuous flux-theory compared to its lattice counterpart.
Consider first the subset of smooth {\it closed} $2d$ surfaces $\tilde{M}$
(resulting from the smooth {\it immersions} of a $2d$ manifold
$M$ into ${\bf R^{D}}$ which alternatively can be represented by the 
maps (\ref{0.1}) with $T\in{\bf R^{D}}$) 
with selfintersections on submanifolds of the dimensionality
$d>d_{cr}=(4-D)$. The latter subset
is of measure zero \cite{StabMap} in the set of all smooth {\it immersions} 
$M\rightarrow{\bf R^{D}}$ (with arbitrary selfintersections). Complementary, 
the smooth immersions $M\rightarrow{\bf R^{D}},~
D\geq{4},$ are dense \cite{StabMap} in the space of all {\it piecewise}
smooth immersions $M\rightarrow{\bf R^{D}}$. In 
particular, in $D\geq{5}$ the subset of $\varphi$-maps resulting in 
smooth embeddings (i.e. in closed
nonselfintersecting $2d$ manifolds
$\tilde{M}$) is the open, dense subspace of the space of all piecewise smooth 
immersions 
$M\rightarrow{\bf R^{D}}$. (Another example, in $D\geq{3}$ the
backtrackings (being viewed as two-dimensional selfintersections) are of
measure zero, i.e. unstable.) 

Next, suppose that the redefinition (discussed in the very end of 
Section 1 and after eq. (\ref{7.11bb})) of the bare $SU(N)$ string 
tension $\tilde{\sigma}_{0}$ is performed that eliminates 
certain types of the collapsed $2d$ subsurfaces originally
built into (\ref{0.1}). Then, by virtue of the Whitney immersion
theorem \cite{StabMap}, the singularities of the generic (piecewise smooth)
mapping (\ref{0.1}) in $D\geq{4}$ can be restricted to the ones listed
after eq. (\ref{0.1}) with the exclusion (in the 
Heat-Kernal case (\ref{4.1b})) of the 
'movable' collapsed subsurfaces attached to a single sheet of the 
covering.  As for the remaining admissible 
collapsed subsurfaces and the branch-points, being necessarily attached
to nontrivial selfintersections of the worldsheet, they therefore
correspond in $D\geq{5}$ to 'measure-zero' 
{\it limiting points} of the dense subspace of the $2d$ manifolds 
$\tilde{M}$ (induced  by the embeddings (\ref{0.1})) {\it without} 
boundaries.
As for $D=4$, the stable (i.e. of nonzero measure) selfintersections of a
closed smooth surface $\tilde{M}$ can occur only at a set of
{\it isolated} points. The advantage of the Gauge String is that, as it
is clear from the previous sections, such zero-dimensional selfintersections
are not weighted by any extra factors so that the corresponding worldsheets
are still assigned with the $J[\tilde{M}(T)|\{\tilde{b}_{k}\}]=1,~b=0,$
reduction of the weight (\ref{0.5c}). Altogether, it justifies the assertion
made in the very end of Section 1.

To see how the $J[\tilde{M}(T)|\{\tilde{b}_{k}\}]\neq{1}$ pattern
is observable in $D\geq{3}$, recall first the basic theorem \cite{StabMap}
on the stability of the smooth immersions: the stability is equivalent to the
local stability. In particular, it implies that the unstable in the case of
a closed  $2d$ surface selfintersections can not be stabilized (in the bulk)
introducing
some selfintersecting boundary contour(s). To be more specific, consider the
framework of the quasiclassical expansion  for $<W_{C}>$ where the weight
of the minimal surface enters as the isolated, discrete contribution disparate
from the continuum of the string fluctuations. As a result of the above
theorem, in $D\geq{4}$ the only place where a nontrivial
$J[\tilde{M}(T)|\{\tilde{b}_{k}\}]\neq{1}$ factor may be observable (beyond the 
considered $SU(N)$ redefinition of $\tilde{\sigma}_{0}$) for a nonbacktracking 
contour $C$
seems to be the contribution of the selfintersecting minimal surface.
The simplest option, where the weights (\ref{0.5cx}) of the movable branch
points can be 'measured', is to take contours
winding a number of times around a nonselfintersecting loop {\it C}.
On the other hand, let {\it C} is any nonbacktracking loop   
associated to some nonselfintersecting  'minimal-area' surface(s).
In this case, we expect
that the simplest $J[\tilde{M}(T)|\{\tilde{b}_{k}\}]=1,~b=1$ pattern 
of (\ref{0.5c}) (with the above redefinition of 
$\tilde{\sigma}_{0}$ and with the worldsheets represented by the smooth strict 
immersions $\tilde{M}:~M\rightarrow{\bf R^{D}}$) in $D\geq{4}$ is 
sufficient to reproduce the correct result for the contribution of the
genus {\it h} smooth worldsheets to the 
average $<W_{C}>$.

\section{Conclusions.}

We propose the correspondence between the smooth Gauge String
(\ref{0.5c}), induced from the lattice models (\ref{4.1}), and the associated
via (\ref{0.1ev}) continuous $YM_{D}$ theory (\ref{0.1ee}) (with a finite
{\it UV} cut off) in
the large {\it N} {\it SC} phase. This duality implies the concrete prediction
(\ref{0.1ef}) for the bare string tension ${\sigma}_{0}=
\Lambda^{2}\tilde{\sigma}_{0}$ as the function of the coupling constants
entering the $YM_{D}$ lagrangian (\ref{0.1ee}). In particular, it readily
allows for a number of nontrivial large {\it N} predictions
in the extreme $SC$ limit where
${\sigma}_{0}$ merges with the leading asymptotics of the physical string
tension ${\sigma}_{ph}$. More generally, the correspondence
asserts that the generic continuous $YM_{D}$ model (\ref{0.1ee}) is confining
in $D\geq{3}$ at least in the large {\it N} {\it SC} regime accessible by our
approach.

In turn, it suggests the mechanism of confinement in the standard weakly
coupled ({\it WC}) $D=4$ continuous gauge theory (\ref{0.1ew}) at large
{\it N}.
Consider the effect of the {\it Wilsonian} renormgroup flow
on the original $YM_{D}$ theory in the {\it WC} phase at the {\it UV} scale
(where the large {\it N} {\it SC} expansion, in terms of the proposed microscopic
gauge strings, fails). The idea is that the latter $YM_{D}$ theory, at the
sufficiently low-energy scale, may be superseded by such effective
strongly coupled $YM_{D}$
system where the (effective) Gauge String representation is already
valid. As the effective $YM_{D}$ system is {\it quasilocal}, we assert that
in the effective Gauge String the Nambu-Goto term in (\ref{0.5c}) is
traded for the
whole
operator expansion (OPE) running in terms of the extrinsic and intrinsic
curvatures of the worldsheet. (Complementary, the weights of the movable
branch points are modified in such a way that the integration, over the
positions of the latter points, results in exactly the same pattern of the
worldsheet OPE as for the descedant of the Nambu-Goto term.)

Finally, let us make contact with the two alternative stringy proposals
\cite{Witt2},\cite{Polyak2}. As for \cite{Witt2}, Witten argues
that in continuous $YM_{4}$ theory (with a finite {\it UV} cut
off $\Lambda$) the physical string tension $\sigma_{ph}$ in the extreme
large {\it N} {\it SC} limit
$g^{2}N\rightarrow{\infty},~g^{2}\sim{O(1/N)},$ scales as
$\sigma_{ph}\sim{g^{2}N\Lambda^{2}}$. On the other
hand, eq. (\ref{0.1ef}) predicts similar large {\it N} $SC$
$O(N\tilde{g}^{2})$-scaling of the bare string tension $\sigma_{0}$
(where $\tilde{g}^{2}\equiv{\tilde{g}^{2}(\{\tilde{b}_{k}\},N)}
\sim{O(N^{-1}\Lambda^{0})}$)
in the following subclass of the $YM_{D}$ systems. The latter
are defined by the subset of the actions (\ref{0.1ee}) with the
coefficients $g_{r}$ constrained by
\be
\tilde{g}^{2}N\rightarrow{\infty}~:~[{g}_{r}]^{-1}
\sim{O(\Lambda^{-2n+4} [N^{\frac{1}{2}}\tilde{g}]^{-n+\gamma(r)}
N^{2-\sum_{k=1}^{n} p_{k}})}~,
\label{7.2n}
\ee
where $\gamma(r)\geq{0}$ (and there is at least one irrep $r$ for which
$\gamma(r)=0$). According to the structure of the (abelianized) Born-Infeld
action, the Witten's prescription \cite{Witt2} is supposed
to induce the $YM_{D}$ action with the local part belonging to the variety
(\ref{7.2n}) (with possible addition of the commutator-terms which does not
alter the conclusions). In sum, our prediction (\ref{0.1ef}) for
the string tension in the extreme large {\it N}  {\it SC} limit
semiquantitatively matches with the $D=4$ pattern of \cite{Witt2} motivated
by the $AdS/CFT$ correspondence.

As for the Polyakov's $D=4$ Ansatz \cite{Polyak2}, it is aimed at a stringy
reformulation of the continuous $YM_{4}$ theory (\ref{0.1ew}) in the
large {\it N} {\it WC} regime. Assuming that the general pattern of the
Ansatz is applicable to the {\it SC} phase as well, an
indirect comparison might be possible. In particular,
the advocated by Polyakov
invariance of the worldsheet action under the extended group of the
diffeomorphisms (with the singularities due to the zero
Jacobian) matches with the two key features of
the Gauge String: the vanishing contributions of the worldsheets $\tilde{M}$
with the backtrackings and the invariance of $w[\tilde{M}]$ under the
group of the area-preserving diffeomorphisms specified in the Section 7.

\begin{center}
{\bf Acknowledgements.}
\end{center}

This project was started when the author was the NATO/NSERC Fellow
at University of British Columbia, and I would like to thank all the staff and
especially Gordon Semenoff for hospitality.

\app{Reconstruction of $\Xi_{4n_{+}}(\{R_{\mu\nu}\})$.}

To derive the $S(4n_{+})$ operator $\Xi_{4n_{+}}(\{R_{\mu\nu}\})$
of eq. (\ref{5.1}), we first rewrite the characters of the original
expression (\ref{5.1bc}) in terms of certain $S(4n_{\mu\nu})$ operator
${\bf D}(\Xi_{4n_{\mu\nu}}(R_{\mu\nu}))$
\be
\chi_{R_{\mu\nu}}(U_{\mu}U_{\nu}U_{\mu}^{+}U_{\nu}^{+})=
Tr_{n_{\mu\nu}}[~{\bf D}(\Xi_{4n_{\mu\nu}}(R_{\mu\nu}))~
{\bf D}_{2}(\{U_{\rho}\otimes U_{\rho}^{+}\})~]~,
\label{BB.1}
\ee
where ${\bf D}_{2}(\{U_{\rho}\otimes U_{\rho}^{+}\})$ is the $D=2$ option of
(\ref{5.1v}). In the
$|\tilde{I}_{4n(\mu\nu)}>$-basis (\ref{5.9}),
${\bf D}(\Xi_{4n_{\mu\nu}}(R_{\mu\nu}))$ reads explicitly
\be
\sum_{\sigma\in{S(n_{\mu\nu})}}
\frac{\chi_{R_{\mu\nu}}(\sigma)}{n_{\mu\nu}!}
\delta^{p_{1}}_{l_{\sigma(3n+1)}}..\delta^{p_{n}}_{l_{\sigma(4n)}}~
\delta^{p_{n+1}}_{l_{1}}..\delta^{p_{2n}}_{l_{n}}~
\delta^{p_{2n+1}}_{l_{n+1}}..\delta^{p_{3n}}_{l_{2n}}~
\delta^{p_{3n+1}}_{l_{2n+1}}..\delta^{p_{4n}}_{l_{3n}}
\label{BB.2}
\ee
which can be represented in the concise form of the
{\it inner}-product
\be
\Xi_{4n_{\mu\nu}}(R_{\mu\nu})=
\Psi_{4n_{\mu\nu}}\cdot \tilde{P}_{4n_{\mu\nu}}(R_{\mu\nu})~~~~;~~~~
\tilde{P}_{4n}(R)={\hat{1}^{\oplus 3}_{[n]}}
\otimes{C_{R}}~,
\label{C.3}
\ee
where $\Psi_{4n_{\mu\nu}}$ is defined by eq. (\ref{5.7}). Each of the four
(ordered) $S(n_{\mu\nu})$-operators in the outer
product composition of $\tilde{P}_{4n_{\mu\nu}}(R_{\mu\nu})$ are postulated to
act on the corresponding four (ordered) $S(n_{\mu\nu})$-subspaces (\ref{5.9})
of the space $|\tilde{I}_{4n(\mu\nu)}>$. Finally, in the $S(4n_{+})$ basis
(\ref{5.6}), one evidently obtains
$\Xi_{4n_{+}}(\{R_{\mu\nu}\})=\otimes_{\{\mu\nu\}}
\Xi_{4n_{\mu\nu}}(R_{\mu\nu})$ that matches with eq. (\ref{5.4}) modulo a
slight deviation of $\tilde{P}_{4n_{\mu\nu}}(R_{\mu\nu})$ from
${P}_{4n_{\mu\nu}}(R_{\mu\nu})$
(defined by eq. (\ref{5.10})).

The possibility to substitute in eq. (\ref{5.1}) the operator
$\tilde{P}_{4n_{\mu\nu}}$ by $P_{4n_{\mu\nu}}$ is ensured by the following
feature of the pattern (\ref{BB.1}). Owing to the basic commutativity
$[{\bf D}(\sigma),U^{\oplus n}]=0$,~
$\forall{\sigma}\in{S(n)}$, any of the unity operators
$\hat{1}_{[n]}$ (the operator $\tilde{P}_{4n}$ is composed
of) can be substituted by the $C_{R}$-factor  properly weighted
according to the multiplication rule $(C_{R})^{2}=C_{R}/d_{R}$. Remark also
that in eq. (\ref{C.3}) the relative order of
the factors $\Psi_{4n_{\mu\nu}}$ and $\tilde{P}_{4n_{\mu\nu}}$ is immaterial
since $[~\Psi_{4n}~,\otimes_{k=1}^{4}\sigma_{k}~]=0,~
\forall{\sigma_{k}}\in{S(n)}$.

Next, let us prove that, in the dual representation
(\ref{5.17}) of (\ref{5.1bc}), the substitution
(\ref{5.21}) (of $\otimes_{\rho=1}^{D}\Phi_{2n_{\rho}}$ by its
square) doesn't change the character (\ref{5.1bd}). The latter 'symmetry' can
be traced back to the fact that the multiple integral (\ref{5.1bc}),
represented by the $S(4n_{+})$ element
(\ref{5.17}), is a real-valued function.
To take advantage of this fact, observe that
$\tilde{\Phi}_{2m}=(\Phi_{2n}\otimes \Phi_{2n}),~
m=2n,$ is the operator
which represents the complex conjugation of the characters 
$\chi_{R}(U_{\mu\nu})=
Tr_{4n}[{\bf D}(\Xi_{4n})
\tilde{U}_{\mu\nu}^{\oplus n}]$
entering (\ref{5.1bc}):
\be
Tr_{4n_{\mu\nu}}[{\bf D}(\Xi_{4n_{\mu\nu}})~
\tilde{U}_{\mu\nu}^{\oplus n_{\mu\nu}}]^{\ast}=
Tr_{4n_{\mu\nu}}[{\bf D}(\Xi_{4n_{\mu\nu}}
\tilde{\Phi}_{2m_{\mu\nu}})~\tilde{U}_{\mu\nu}^{\oplus n_{\mu\nu}}],
\label{C.11}
\ee
where $\Xi_{4n_{\mu\nu}}
\equiv{\Xi_{4n_{\mu\nu}}(R_{\mu\nu})}$ is defined in eq. (\ref{C.3}), and
$\tilde{U}_{\mu\nu}$ is introduced in eq. (\ref{5.21bb}).
As the master-integral (\ref{5.1bc}) is invariant under the simultaneous
transformation (\ref{C.11}) of all the involved characters while
$\otimes_{\mu\nu} \tilde{\Phi}_{2m_{\mu\nu}}=
\otimes_{\rho=1}^{D} \Phi_{2n_{\rho}}$,
we arive at the required invariance under (\ref{5.21}).

As for eq. (\ref{C.11}), it can be deduced by linearity from the following
more elementary identity. To formulate the latter, let us first introduce
the representation of the basic traces (the characters
(\ref{C.11}) are composed of)
\be
tr({(U_{\mu}U_{\nu}U_{\mu}^{+}U_{\nu}^{+})^{n}})\equiv
{Tr_{4n}[{\bf D}(\Gamma_{4n}
\Psi_{4n})\tilde{U}_{\mu\nu}^{\oplus n}]}=
{Tr_{4n}[{\bf D}(\Gamma^{-1}_{4n}
\Psi_{4n})\tilde{U}_{\mu\nu}^{\oplus n}]},
\label{C.9}
\ee
where $\Gamma_{4n}=(c_{n}\otimes{\hat{1}^{\oplus 3}_{[n]}})$ in the $S(4n)$
basis (\ref{5.9}), and $c_{n}$ is the {\it n}-cycle permutation.
Then, the required identity reads (with $m=2n$)
\be
Tr_{4n}[{\bf D}(\Gamma_{4n}~
\Psi_{4n})~\tilde{U}_{\mu\nu}^{\oplus n}]^{\ast}=
Tr_{4n}[{\bf D}(\Gamma_{4n}~
\Psi_{4n}~\tilde{\Phi}_{2m})~\tilde{U}_{\mu\nu}^{\oplus n}].
\label{C.9bb}
\ee
For its justification. we first introduce the tensor representation of the
relevant complex conjugation:
$Tr_{n}[{\bf D}(\sigma)\otimes_{k=1}^{n}U_{k}]^{\ast}=
Tr_{n}[{\bf D}(\sigma^{-1})\otimes_{k=1}^{n}U^{+}_{k}]$ where the orderings of
the $U_{k}$ factors in the l.h.s. is the same as that of $U^{+}_{k}$ in the
r.h.side. The key-observation is that, when
$\otimes_{k=1}^{n}U^{+}_{k}=\tilde{U}_{\mu\nu}^{\oplus n_{\mu\nu}}$, the
substitution $U_{k}\rightarrow{U^{+}_{k}}$ can be performed
as the conjugation with respect to
$(\Phi_{2n_{\mu\nu}}\otimes \Phi_{2n_{\mu\nu}})=\tilde{\Phi}_{2m_{\mu\nu}}=
\tilde{\Phi}^{-1}_{2m_{\mu\nu}}$:
$\tilde{\Phi}^{-1}_{2m_{\mu\nu}}\cdot \tilde{U}_{\mu\nu}^{\oplus n_{\mu\nu}}
\cdot \tilde{\Phi}_{2m_{\mu\nu}}=\tilde{V}_{\mu\nu}^{\oplus n_{\mu\nu}}$
where $\tilde{V}_{\mu\nu}=
U^{+}_{\mu}\otimes U_{\mu}\otimes U^{+}_{\nu}\otimes U_{\nu}$. This is 
because, owing to eq. (\ref{5.15}), $\Phi_{2n}$ interchanges
(either upper
or lower) indices between the $U^{\oplus n}$ and
$(U^{+})^{\oplus n}$ blocks of $U^{\oplus n}\otimes (U^{+})^{\oplus n}$. As
in  eq. (\ref{C.9}) $\Gamma_{4n}$ can be substituted by $\Gamma^{-1}_{4n}$
(while $\Gamma_{4n}$ commutes with both $\tilde{\Phi}_{2m}$ and
$\Psi_{4n}$), all what remains to be proved is that
$\Psi_{4n}\tilde{\Phi}_{2m}=\tilde{\Phi}_{2m}\Psi_{4n}=\Psi^{-1}_{4n},~n=2m$.
The last identity, owing to the specific patterns (\ref{5.15}) and (\ref{5.7})
of ${\Phi}_{2n}$ and $\Psi_{4n}$, directly follows from its reduced $m=2$
variant. This completes the justification of (\ref{C.9bb}).

Finally, let us sketch  the proof of the basis formula (\ref{5.17b}). To make
sure that the contraction of the $\sigma^{(\pm)}_{\rho}$ indices is the same
in the both sides of (\ref{5.17b}),
the following decomposition of the index-structure is helpful. Building on the
prescription (\ref{3.18}), we rewrite the $S(n_{\rho})$ operators
$D(\sigma^{(\pm)}_{\rho})$ in the form
\be
D(\sigma^{(\pm)}_{\rho})^{\{j^{\oplus n_{\rho}}\}}
_{\{i^{\oplus n_{\rho}}\}}\equiv{D}(\sigma^{(\pm)}_{\rho})
^{\{j^{\oplus n_{\rho\mu}}\}...\{j^{\oplus n_{\rho\nu}}\}}
_{\{i^{\oplus n_{\rho\mu}}\}...\{i^{\oplus n_{\rho\nu}}\}}~~,~~
\mu,...,\nu{\neq{\rho}},
\label{C.15}
\ee
where each of the $D-1$ $n_{\rho\lambda}$-subsets of the indices
($\lambda{\neq{\rho}}$) acts on the associated $S(n_{\rho\lambda})$ subspace
(\ref{5.6c}) of the
enveloping space $S(n_{+})$. Let consider any $n_{\rho\nu}$-subset,
associated to $\sigma^{(\pm)}_{\rho}$, as the composite block-index
$j^{(\pm)}_{\rho\nu}$. A direct inspection
then reveals that, in {\it both} sides of eq. (\ref{5.17b}), for a given
$\mu,\nu$ the four block-indices $j^{(\pm)}_{\mu\nu},j^{(\pm)}_{\nu\mu}$
(corresponding to either $\rho=\mu$ or $\rho=\nu$) are $c_{4}$-cyclically
contracted according to the pattern (\ref{5.4})/(\ref{5.7}) of
$\otimes_{\mu\nu}\Psi_{\mu\nu}$.

As for the ordering of the inner $\rho$-products in the r.h.s. of
(\ref{5.17b}), it should be deduced from the particular
ordering of the $|i_{\pm}(\rho)>$-blocks composed into the
elementary subspace (\ref{5.7b}) used to define the ${D(D-1)/2}$ operators
$\Psi_{4n_{\mu\nu}}$. To justify the prescription stated in Section 3,
one is to combine the above analysis with the specified pattern of the
ordering in the elementary $D=2$ case (\ref{5.21b}).

\app{Tensor representation vs. Regular one. }

Let us first derive the identity (\ref{5.1dd}). Actually, this
equation is nothing but the transformation of the trace of the tensor
${\bf D}(\sigma)$-representation (\ref{3.7}) into that of the canonical
regular representation $X_{REG}$ \cite{Gr-in-phys} (both
associated to a given $S(n)$-algebra). Recall that $X_{REG}$ is defined
\cite{Gr-in-phys} on the vector space $\Theta=\{\sigma_{i};~
\sigma_{i}\in{S(n)}\}$ by the homomorphism $S(n)\rightarrow
{M_{i}\in{End(\Theta)}}$ of the $S(n)$-group into the group of the
endomorphisms of $\Theta$: $\sigma_{i}\sigma_{j}=\sigma_{k}\Delta^{k}_{ij}
\rightarrow (M(\sigma_{i}))^{k}_{j}\equiv{(M_{i})^{k}_{j}}=\Delta^{k}_{ij}$.
Here $\sigma_{i}\sigma_{j}=\sigma_{q}$, and $\Delta^{k}_{ij}=1$ or $0$
depending on whether $q=k$ or $q\neq{k}$. Defined in this way, the matrices
$M_{i}$ satisfy the same relations
$M_{i}M_{j}=M_{q}$ as the origibal $S(n)$ group-elements $\sigma_{i}$.

To make contact with the tensor representation (\ref{3.7}), recall
\cite{Gr-in-phys} first that
\be
\chi_{REG}(\sigma)=\sum_{R\in{Y_{n}}} d_{R}~\chi_{R}(\sigma)=
n!~\delta_{n}(\sigma)~~,~~\chi_{REG}(P_{R}\sigma)=d_{R}\chi_{R}(\sigma),
\label{AA.1}
\ee
where $tr[M(\sigma)]\equiv{\chi_{REG}(\sigma)}$ and $P_{R}=d_{R}C_{R}$. As
for $\delta_{n}(..)$,
on the $S(n)$-group it reduces to the standard Kronecker $\delta$-function:
$\delta_{n}(\sigma)=\delta[{\sigma,\hat{1}_{[n]}}]$ (with
$\hat{1}_{[n]}$ being the 'trivial' unity-permutation of $S(n)$).
By linearity, it is then extended to the $S(n)$-algebra. 
Finally, rewriting the $V=\hat{1}$ reduction of the second Frobenius formula
(see e.g. \cite{Moore,Dub2}) in terms of $\chi_{REG}(C_{R}\sigma)$
\be
Tr_{n} [{\bf D}(\sigma)]=
\sum_{R\in{Y_{n}^{(N)}}} \chi_{R}(\sigma)\chi_{R}(V)|_{V=\hat{1}}=
\sum_{R\in{Y_n^{(N)}}} dimR~\chi_{REG}(C_{R}\sigma),
\label{A.12}
\ee
and employing the definition (\ref{3.8}) of $\Lambda^{(1)}_{n}$ (where the
sum runs over the chiral {\it U(N)} irreps {\it R}), we arrive at the basic
identity (\ref{5.1dd}). (Omitting the $P_{n}^{(N)}$ projector, the latter
identity was discussed in \cite{Ramg}.)

In conclusion, let us briefly sketch the derivation of eqs. (\ref{7.1}),
(\ref{7.5}). As for (\ref{7.1}), in its l.h.s. we use first (\ref{7.4}) to
substitute $d_{R}(n!dimR/d_{R})^{m}$ by $\chi_{R}((N^{n}\Omega_{n})^{m})$.
Replacing $C_{R}=\sum_{\sigma\in{S(n)}}\chi_{R}(\sigma^{-1})\sigma/n!$,
we then combine the two resulting characters into one
\be
\frac{1}{n!}\sum_{\sigma\in{S(n)}}\sum_{R\in{Y_{n}}} d_{R}~
\chi_{R}((N^{n}\Omega_{n})^{m}P_{n}^{(N)}\sigma^{-1})~{\bf D}(\sigma)~.
\label{BB.10}
\ee
As $[\Omega_{n},\sigma]=0,~\forall{\sigma}\in{S(n)}$, for the derivation of
(\ref{BB.10}) we used the identity $\chi_{R}(\Psi\sigma)=
\chi_{R}(\sigma)\chi_{R}(\Psi)/d_{R},~
\forall{\rho\in{S(n)}}$, if $[\Psi,\rho]=0$ (which directly follows from the
possibility \cite{Gr-in-phys} to expand any
such $\Psi$, in the center of $S(n)$, in terms of the Young projectors
$\Psi=\sum_{R\in{Y_{n}}} \psi_{R}P_{R}$). We have used also that
$\chi_{R}(P_{n}^{(N)}\sigma)=\chi_{R}(\sigma)$ or $0$ depending on whether or
not $R\in{Y_{n}^{(N)}}$. Applying the completeness condition
$\sum_{\sigma\in{S(n)}}\delta_{n}(\sigma^{-1}\Phi)~
{\bf D}(\sigma)={\bf D}(\Phi)$ (where $\Phi$ is {\it any} element of
the $S(n)$-algebra) we arrive at the announced result (\ref{7.1}).

Concerning eq. (\ref{7.5}), first one is to represent (see e.g.
\cite{Gr&Tayl}): $C_{2}(R)=(nN+
2\chi_{R}(\hat{T}^{(n)}_{2})/d_{r}-n^{2}/N),~R\in{Y_{n}},$ which is a
particular case of the general relation (\ref{0.7vv}). Expanding
$\chi_{R}(\hat{T}^{(n)}_{2})/d_{r}$ (where $\hat{T}^{(n)}_{2}$ is defined by
eq. (\ref{7.6})) into the preexponent, the derivation
of (\ref{7.5}) is given by a minor modification of the steps developed in the
context of (\ref{7.1}).

\app{Reintroducing the branch points.}

Given the generalized covering spaces (GCSs) $\tilde{M}(\{\sigma_{\rho}\})$ of
$T_{EK}$ (resulting after the deformation  (\ref{7.17cb}) of (\ref{7.7})), the
generalized branched covering spaces (GBCSs) of $T_{EK}$ (encoded in the
full expression (\ref{7.17})) can be reconstructed reintroducing onto
$\tilde{M}(\{\sigma_{\rho}\})$ the relevant branch points (BPs).
Alternatively,
the GBCSs of $T_{EK}$ are reinterpreted as the GCSs of $T_{EK}-\{q_{s}\}$
where the deleted (from $T_{EK}$) set of points $\{q_{s}\}$ is
associated to the corresponding BP permutations $\xi^{\{p_{\phi}\}},~
\tau^{(k_{\mu\nu})}_{\mu\nu}$ in compliance with the homomorphism
(\ref{7.12cb}). Then, a particular GBCS of $T_{EK}$ is reproduced from the
associated GBC (of $T_{EK}-\{q_{s}\}$) 'closing' the temporarily
deleted points $\{\varphi^{-1}(q_{s})\}$. At a given $q_{s}$, the closure
\cite{Ezell,Moore} is performed as the mapping of
the $n_{\phi(s)}$-set $\varphi^{-1}(q_{s})\cong{\Upsilon_{\phi(s)}}$ onto
set of points matching with the number of cycles in
the cyclic decomposition of the associated to $q_{s}$
permutation (\ref{7.12cb}).

To make the above procedure well-defined, I propose to view $T_{EK}$ as the
{\it homotopy retract} of a 'less singular' $2d$ cell-complex $T'_{EK}$ with
the same fundamental group: $\pi_{1}(T'_{EK})=\pi_{1}(T_{EK})$. 
As the proposed retraction preserves the basic homomorphism (\ref{7.12cb}),
the GBCSs of $T_{EK}$ can be consistently treated as the limiting case of the
corresponding GBCSs of $T'_{EK}$.

First, let us
thicken each $\rho$-link into an infinitesimally thing 
cylinder $\bar{Z}_{\rho}=\cap_{\nu\neq{\rho}} E_{\rho\nu}$ shared by the $D-1$
corresponding 2-tora $E_{\rho\nu}$. (To match with the canonical
construction \cite{Ezell,Moore} of the branch points, (for a given $\rho$)
all $E_{\rho\nu},~\nu\neq{\rho},$ can be always
adjusted to have the {\it same} orientation when restricted to
$\bar{Z}_{\rho}$.) Complementary, the intersection
$\bar{Z}_{+}=\cap_{\rho} \bar{Z}_{\rho}=\cup_{\rho\nu} E_{\rho\nu}$ of the $D$
different cylinders $\bar{Z}_{\rho}$ (or, equivalently, of all the $D(D-1)/2$
$\mu\nu$-tora $E_{\mu\nu}$) is thicken into an infinitesimal disc
$\bar{Z}_{+}$. (Again, all the cylinders $\bar{Z}_{\rho}$ can be adjusted 
to have the {\it same} orientation when restricted to $\bar{Z}_{+}$.)

Choose a base-point $p\neq{q_{s}},~\forall{s,}$ (common for all the
equivalence classes of the pathes represented in the constraint
(\ref{7.12ccc})) in the interior $Z_{+}$ of $\bar{Z}_{+}$.
We require that the location of the set
$\{q_{s}\}=\cup_{\phi}\{q^{(\phi)}_{k_{\phi}}\}$ complies with
\be
q^{(\mu\nu)}_{k_{\mu\nu}}\in{(E_{\mu\nu}-(\bar{Z}_{\mu}\cup \bar{Z}_{\nu}))}
~~;~~q^{(\rho)}_{k_{\rho}}\in{(Z_{\rho}-\bar{Z}_{+})}~~;~~
q^{(+)}_{k_{+}}\in{Z_{+}}~,
\label{7.16b}
\ee
where $Z_{\phi}$ is the {\it interior} of the closed space $\bar{Z}_{\phi},~
\phi\in{\{\mu\nu\},\{\rho\},+}$. Also, the introduced in Section 6 closed
branch cuts $\varpi_{\rho}$ (inherited from the GCS of $T'_{EK}$) are supposed
to satisfy: $\varpi_{\rho}\in{Z_{\rho}}$, while
$\{q^{(\phi)}_{k_{\phi}}\}\cap \varpi_{\rho}=0,~\forall{\rho},~
\forall{k_{\phi}}$.

To reintroduce the branch points onto the GCSs $\tilde{M}(\{\sigma_{\rho}\})$
of $T'_{EK}$, one is to consider the additional (to
$\varpi_{\rho}$) branch cuts $\varpi_{\phi(s)}$ outgoing from the
corresponding points $q_{s}$ of $T'_{EK}$. (Combining all the
cuts together, one is supposed to arrive at a network
$\Omega=\{\varpi_{\phi(s)},\varpi_{\rho}\}$ whose global consistency is
ensured by the $\delta_{n_{+}}$-constraint (\ref{7.17}).) To implement
$\varpi_{\phi(s)}$, we cut the GCS $\tilde{M}(\{\sigma_{\rho}\})$
along the $\varphi^{-1}$-image of $\varpi_{\phi(s)}$ (starting at
$q_{s}$ and terminating either at some other point $q_{k}$ or, possibly, at
an auxiliary vertex of the network $\Omega$). Denote
$\alpha(s_{k})$ and $\beta(s_{k})$ (with $k=1,...,n_{\phi(s)}$) the resulting
$n_{\phi(s)}$ boundaries of the sheets (of $\tilde{M}(\{\sigma_{\rho}\})$) 
respectively on the left and on the right sides of $\varpi_{\phi(s)}$.
To obtain the GCS of $T'_{EK}-\{q_{s}\}$ corresponding to the
homomorphism (\ref{7.12cb}), one is to perform the pairwise
identifications
of these boundaries according to the developed prescription (\ref{7.16}).
Matching with (\ref{7.17}), one
simply substitutes in (\ref{7.16}): $\rho\rightarrow{s}$. so that
$\sigma_{s}$ is either $\xi^{\{p_{\phi}\}}$ or
$\tau^{(k_{\mu\nu})}_{\mu\nu}$ depending on the image of the
homomorphism (\ref{7.12cb}) assigned to a given branch point $q_{s}$ constrained by
(\ref{7.16b}).

\app{Counting the generalized coverings.}

The proof, that inequivalent spaces $\tilde{M}_{\varphi}$ in (\ref{7.17})
are parametrized by the equivalence classes (\ref{7.12ex}) of the
homomorphisms (\ref{7.12cb}), appropriately generalizes the analogous proof
\cite{Massey,Ezell} for the canonical coverings employed in
\cite{Gr&Tayl,Moore}. First, one is to demonstrate that the relevant
inequivalent spaces
$\tilde{M}_{\varphi}$ entering (\ref{7.17}) are uniquely parametrized
by certain $\otimes_{\mu\nu} S(n_{\mu\nu})$
conjugacy classes of the $\pi_{1}(T'_{EK}-\{q_{s}\}|p)$ subgroups.
Then, one proves the one-to-one correspondence between the latter classes of
the subgroups and the equivalence classes (\ref{7.12ex}) of the
homomorphisms (\ref{7.12cb}). Let us outline the above proofs with the
emphasis on the subtleties novel compared to \cite{Massey,Ezell}.

To begin with, we observe that (by the same token \cite{Massey} as in the
canonical case) any subgroup of $\pi_{1}(T'_{EK}-\{q_{s}\}|p)$ can be
viewed as the image
\be
\varphi_{\ast}(\pi_{1}(\tilde{M}_{\varphi}-\{\varphi^{-1}(q_{s})\}|
\tilde{p}))~~~,~~~\tilde{p}\in{\varphi^{-1}(p)}\cong{\Upsilon_{n_{+}}}~,
\label{7.17vxv}
\ee
induced by the $\varphi$-mapping (\ref{0.1}): $\tilde{M}_{\varphi}
\rightarrow{T'_{EK}}$, reconstructed in Section 6.1 and Appendix C.
The consistency of (\ref{7.17vxv}) implies that a given $\tilde{M}_{\varphi}$
yields some branched covering (with $n_{+}$-sheets) 'above' the single site
{\it s} of $T_{EK}$ which is 'regularized', $s\rightarrow{Z_{+}}$ (see eq.
(\ref{7.16b}) of Appendix C), into an infinitesimal disc $Z_{+}$ of $T'_{EK}$
so that $p\in{Z_{+}}$. (In contradistinction to the canonical case, the
point {\it p} is {\it not} allowed to leave $Z_{+}$.)  Perform a shift of the
base-point $\tilde{p}\in{\varphi^{-1}(p)}$, located on a $j$th sheet of the
branched covering of $Z_{+}$, to $\tilde{p}'\in{\varphi^{-1}(p)}$
on some other $k$th sheet along the path $\varepsilon
\in{\tilde{M}_{\varphi}-\{\varphi^{-1}(q_{s})\}}$. It induces \cite{Massey}
the associated conjugation of the $\pi_{1}(T'_{EK}-\{q_{s}\}|p)$ elements
(and, hence, its subgroups) with respect to the element given by the image
$\varphi(\varepsilon)\in{\pi_{1}(T'_{EK}-\{q_{s}\}|p)}$ of the path
$\varepsilon$.

Next, one notes that (similarly to the canonical construction
\cite{Massey}) the shift $\varepsilon$ acts on the $n_{+}$-set
$\varphi^{-1}(p)\cong{\Upsilon_{n_{+}}}$ as the simple
$S(n_{+})$-transposition
$\psi(\varphi(\varepsilon))\in{T^{(n_{+})}_{2}}$ of the two corresponding
entries: $j\rightarrow{k},~j,k,=1,...,n_{+}$. As a result,
$\varepsilon$ induces the $S(n_{+})$ conjugation (\ref{7.12ex})
of (\ref{7.12cb}) with $\eta=\psi(\varphi(\varepsilon))$ that can be
visualized as the permutation of the two associated sheets of
$\tilde{M}_{\varphi}$. The latter does not necessarily
leaves intact the topology of $\tilde{M}_{\varphi}$. The cutting-gluing
technique of Section 6.1 implies that
interchanges of the sheets encoded in the conjugations with
$\eta\in{S(n_{+})/\otimes_{\mu\nu}S(n_{\mu\nu})}$ (i.e. between the
$n_{\mu\nu}$-sheet coverings of different $\mu\nu$-tora $E_{\mu\nu}$ of
$T'_{EK}$) should be excluded when collecting an equivalence class of maps
(\ref{7.12ex}) corresponding to a given topology of $\tilde{M}_{\varphi}$.

Consider the $\otimes_{\mu\nu} S(n_{\mu\nu})$ conjugacy classes of the
$\pi_{1}(T'_{EK}-\{q_{s}\}|p)$ subgroups induced by the combination of the
elements $\varphi_{\ast}(\varepsilon)$ which are the images of those of the
shifts $\varepsilon$ that connect the sheets within the $n_{\mu\nu}$-sheet
covering of a given torus
$E_{\mu\nu}$ of $T'_{EK}$. Taking into account the above discussion, one
readily modifies the canonical construction \cite{Massey} to
prove that the considered classes
of the subgroups are indeed in the 1-to-1 correspondence with
the inequivalent generalized covering spaces
$\tilde{M}_{\varphi}-\{\varphi^{-1}(q_{s})\}$ of
$T'_{EK}-\{q_{s}\}$ (and, therefore, with inequivalent GBCSs
$\tilde{M}_{\varphi}$ of $T'_{EK}$). In particular, the cutting-gluing rules
of Section 6.1 ensure that the mapping from $\tilde{M}_{\varphi}$ to the
corresponding classes of the $\pi_{1}(T'_{EK}-\{q_{s}\}|p)$ subgroups is onto.

Finally, let us reveal the second 1-to-1 correspondence of the former
classes of the subgroups with the equivalence classes (\ref{7.12ex}) of the
homomorphisms (\ref{7.12cb}). From above, it is clear that
the $\psi$-images (\ref{7.12cb}) of these $\pi_{1}(T'_{EK}-\{q_{s}\}|p)$
subgroups can be related by the $\otimes_{\mu\nu} S(n_{\mu\nu})$ permutations
corresponding to the interchanges of the sheets separately
within each of the $D(D-1)/2$ $n_{\mu\nu}$-sheet coverings associated to
particular $E_{\mu\nu}$-tora of $T'_{EK}$. As a result, akin to the
$D=2$ case \cite{Massey,Ezell}, the latter permutations are represented by
the required conjugations (\ref{7.12ex}) with
$\eta\in{\otimes_{\mu\nu} S(n_{\mu\nu})}$. Conversly, equivalent
homomorphisms determine equivalent $\otimes_{\mu\nu} S(n_{\mu\nu})$ classes of
the $\pi_{1}(T'_{EK}-\{q_{s}\}|p)$ subgroups (according to the discussed above
construction of $\eta=\psi(\varphi(\varepsilon))$).
Summarizing, it completes the proof of the two asserted correspondences.

\app{The higher Casimirs' actions.}

To derive the generalized form (\ref{7.6b}),(\ref{7.6c}) of 
$Q_{n}(\Gamma)$ in  (\ref{7.5}), first one is to
begin with the (Schur-Weyl) duality \cite{Moore,Ganor} between the Casimir
eigenvalues
\be
\frac{C_{q}(R)}{N^{q-1}}=
\sum_{{T}_{\{p\}}\in{S(n)}} a_{q}(N,n,T_{\{p\}})~
\frac{\chi_{R}(\hat{T}_{\{p\}})}{d_{R}}~~~,~~~R\in{Y_{n}^{(N)}}~,
\label{0.7vv}
\ee
and the symmetric group characters $\{\chi_{R}(\hat{T}_{\{p\}})\}$
(where $\hat{T}^{(n)}_{\{p\}}\equiv{\hat{T}_{\{p\}}}$ is defined by eq.
(\ref{7.6b}), while $1\leq{q}\leq{N}$). The coefficients $a_{q}(...)$ are
defined by the formular
\cite{Moore,Ganor} which can be deduced from eq. (\ref{7.6c}) via the substitution:
$\upsilon_{\{p\}}(..,n,N)\rightarrow{a_{q}(N,n,T_{\{p\}})},~
s_{\{p\}}(\{\tilde{b}_{r}\},m,l)
\rightarrow{s_{q}(T_{\{p\}},m,l)},~M_{\{p\}}
\rightarrow{M_{0}(T_{\{p\}})}$. In particular, (keeping $n\sim{O(N^{0})}$) the
{\it leading} term of the formal $1/N$ expansion
of the $p$th order Casimir operator reads: $C_{q}(R)=
N^{q-1}(n+O(N^{-1})),~R\in{Y_{n}^{(N)}}$, which
corresponds in (\ref{0.7vv}) to the trivial unity permutation
$T_{\{p\}}=\hat{1}_{[n]}$. As for
the remaining $T_{\{p\}}\neq{\hat{1}_{[n]}}$ contributions, the branch point
interpretation of the $l=0$ term in expression like  (\ref{7.6c}) is
discussed in the end of
Section 5 (in fact $s_{q}(T_{\{p\}},m,0)=0$ for $m\geq{2}$). To interprete a
given $l\geq{1}$ term, first one is to resum the powers
$n^{m}$ in terms of the numbers of
inequivalent subdivisions of {\it n} objects into two subsets containing $t$ and
$(n-t)$ objects: $n^{m}=\sum_{t=1}^{m} f_{m,n} n!/t!(n-t)!$. As a result,
similarly to \cite{Ganor,Moore} the latter contributions include 
(in addition to the ${T}_{\{p\}}$ branch point) the extra
collapsed to a point $2d$ subsurfaces. Being cut out, these surfaces can be
viewed as having genus $g=l-[t/2]$ and $2[t/2]$ holes.

Finally, 
to select the admissible class of the functions $\Gamma$ (defining a given
model (\ref{4.1})), one is to require that in the associated large $N$
{\it SC} series (\ref{0.5}) the factor $N^{2-2h}$ matches with the genus $h$
of the associated worldsheet $\tilde{M}$. Then, to ensure eq. (\ref{7.6c}),
the admissible polynomial pattern of the function $\Gamma$
should  belong to the variety (with $M\in{\bf Z_{\geq{1}}}$)  
\be
\Gamma(\{\tilde{b}_{r}\},N,\{C_{q}(R)\})=\sum_{M,\{q_{k}\}}
\sum_{\bar{l}\geq{[\frac{M}{2}]}}
s(\{\tilde{b}_{r}\},\{q_{k}\},\bar{l})~ N^{-2\bar{l}} \prod_{k=1}^{M}
\frac{C_{q_{k}}(R)}{N^{q_{k}-1}},
\label{0.7vvv}
\ee
where the (properly weighted by the {\it 1/N} factors) $\bar{l}\geq{1}$ terms
encode the additional, compared to those encoded in (\ref{0.7vv}), collapsed
subsurfaces connecting a few sheets according to the same rules as for
(\ref{0.7vv}). In particular, the pattern (\ref{0.7vvv})/(\ref{0.7vv})
ensures the existence of the proper 'asymptotics' (\ref{0.1ef}) of 
$\Gamma(...)$ defining the bare string tension $\sigma_{0}(\{\tilde{b}_{r}\})$
(to which only {\it linear} in $C_{q}(R)$ terms in (\ref{0.7vvv}) contribute).

Finally, with the help of some
elementary identities \cite{Gr&Tayl} from the theory of
$\chi_{R}(\hat{T}_{\{p\}})$ characters, the generic function
(\ref{0.7vvv}) results in the generic pattern (\ref{7.6b}) of the
generalized operator $Q_{n}(\Gamma)$.
One can argue \cite{Ganor} that the required pattern (\ref{7.6c}) of the
{\it N}-dependence takes place provided, in the standard
$F_{\mu\nu}$-representation (\ref{0.1ee}), the $1/N$ scaling of the
coefficients $g_{r}$ is constrained to yield the conventional 't Hooft
pattern of the $1/N$ topological expansion.

\end{document}